\documentclass[iop,apj,12pt,numberedappendix]{emulateapj}
\usepackage{times}
\usepackage[normalem]{ulem}
\usepackage{graphicx}
\usepackage{natbib}
\usepackage{amsfonts}
\usepackage{amsmath}
\usepackage{multirow}
\usepackage{enumerate}
\usepackage{comment}
\usepackage{float}
\usepackage{verbatim}
\usepackage[para]{threeparttable}
\usepackage[usenames,dvipsnames]{xcolor}
\usepackage[plainpages=false, colorlinks=true, anchorcolor=blue, linkcolor=blue, citecolor=blue, bookmarks=false]{hyperref}
\usepackage{scrextend}
\newcommand{\be}{\begin{eqnarray}}
\newcommand{\ee}{\end{eqnarray}}

\newcommand{\shortauth}{Morozova et al.}
\newcommand{\slugcom}{Accepted for publication in The Astrophysical Journal}
\slugcomment{\slugcom}

\newcommand{\GG}[1]{}

\lefthead{\sc \footnotesize \slugcom \hfill \shortauth}
\righthead{\sc \footnotesize \slugcom \hfill \shortauth}

\begin{document}

\title{Measuring the Progenitor Masses and Dense Circumstellar Material of Type II Supernovae}

\author{Viktoriya Morozova\altaffilmark{1}}
\author{Anthony L. Piro\altaffilmark{2}}
\author{Stefano Valenti\altaffilmark{3}}
\altaffiltext{1}{Department of Astrophysical Sciences,
  Princeton University, Princeton, NJ 08544, USA;
  vsg@astro.princeton.edu}
\altaffiltext{2}{The Observatories of the Carnegie Institution for Science, 813 Santa Barbara St., Pasadena, CA 91101}
\altaffiltext{3}{Department of Physics, University of California, Davis, CA 95616, USA}

\begin{abstract}

Recent modeling of hydrogen-rich Type II supernova (SN II) light curves suggests
the presence of dense circumstellar
material (CSM) surrounding the exploding progenitor stars. This has important implications for the
activity and structure of massive stars near the end of their lives. Since previous work
focused on just a few events, here we expand to a larger sample of
twenty well-observed SNe II. For each event we are able to constrain the
progenitor zero-age main-sequence (ZAMS) mass,
 explosion energy, and the mass and radial extent of dense CSM.
We then study the distribution of each of these properties across the full sample of SNe.
The inferred ZAMS masses are found to be largely consistent with
a Salpeter distribution with minimum and maximum masses of $10.4$ and $22.9\,M_{\odot}$,
respectively.  We also compare
the individual ZAMS masses we measure with specific SNe II that have pre-explosion
imaging to check their consistency. Our masses are generally comparable to or larger than
the pre-explosion imaging masses, potentially helping ease the red supergiant problem.
The explosion energies vary from
$(0.1-1.3)\times10^{51}\,{\rm erg}$, and for $\sim70\%$ of the SNe we obtain CSM masses
in the range between $0.18-0.83\,M_{\odot}$. We see a potential 
correlation between the CSM mass and  explosion energy, which  suggests
that pre-explosion activity has a strong impact on the structure of the star. This may be important to
take into account in future studies of the ability of the neutrino mechanism to explode stars.  We also see a possible
correlation between the CSM's radial extent and ZAMS mass, which could be related
to the time with respect to explosion when the CSM is first generated.

\end{abstract}

	
  
\section{Introduction}
\label{sec:introduction}

A longstanding problem in the study of explosive transient events is
connecting classes of supernovae (SNe) to the specific progenitors
that generate them. In this sense, the hydrogen-rich Type II SNe should
seemingly be the most straightforward for making this connection.
With the exception of the rarer Type IIb and 1987A-like events, pre-explosion
imaging
obtained with the {\em Hubble Space Telescope} exclusively identifies
red supergiants (RSGs) as their progenitors \citep{li:06,smartt:09a,vandyk:12}.
These are expected to largely evolve as single stars, so that their
models can be generated without many of the complications present for
other stellar calculations (such as binarity, a high spin, etc). This in turn
should simplify explosion and light curve calculations, which find that a RSG naturally
produces the plateau-shaped light curve of SNe II
with the recombination of many solar masses of
hydrogen-rich material with a RSG-like radius
\citep{smith:11,utrobin:13,dessart:13,gonzalez:15,rubin:16,renzo:17}.

Nevertheless, there remain many outstanding questions about how Type II
progenitors connect to their light curves. Chief among these is
the relationship between the Type IIP (plateau) and Type IIL (linear) subclasses,
which is based on the shape of their light curves during the first few weeks
\citep{barbon:79}. There has been a long debate on whether there is a physical
variable that smoothly transitions between Type IIP and IIL or whether there
is a specific mechanism that creates this dichotomy more abruptly. There have
 been some claims of distinct populations \citep{arcavi:12,faran:14a,faran:14b},
 but support for the more continuous case has increased as larger compilations by
\citet{anderson:14} and \citet{sanders:15} showed a more continuous range
 of early light-curve slopes. An important breakthrough came when
 \citet{valenti:15} demonstrated that if
 one simply follows an SN IIL long enough, its light curve will drop at $\sim100$ days,
 just like a normal SN IIP (previous SNe IIL studies rarely followed the light curve
 beyond $\sim80$ days from discovery; see also \citealp{anderson:14}). This suggests
 that Type IIL and Type IIP SNe may have a similar amount of hydrogen present
 for the main bulk of their envelopes, and whatever is creating the Type IIL distinction
 may be contributing something above a fairly normal underlying RSG.

Motivated by these issues, \citet{morozova:17} recently numerically modeled
three Type II SNe with extensive photometry over many wavebands by using RSG
models with dense CSM stitched on top of them. This work was different from previous
theoretical studies \citep[e.g.,][]{moriya:11} in that the CSM was generally more
massive and compact (only extending a few stellar radii above the RSG). The two
important conclusions from this work were that (i) this dense CSM could naturally explain the
photometric differences between the Type IIP and IIL, and (ii) even the seemingly
more normal Type IIP SNe need dense CSM to accurately model their light curves.
It is still unclear what the full implications of these results are. It seems to indicate
that there is increased activity in RSGs during the last months or years of their
lives, which may be related to theoretical studies of pre-explosion outbursts
\citep{quataert:12,shiode:14,quataert:16,fuller:17}. Furthermore, observations
of SNe II shortly after explosion show narrow lines that indicate a dense wind-like
environment \citep[e.g.,][]{yaron:17}, albeit probing more extended and less dense
material than the CSM we need for the light curves \citep{dessart:17}. This in turn brings up the question
of whether there is a relationship between these two components of the CSM
\citep{moriya:17}.

An important step toward better understanding this dense CSM is mapping
out its diversity over a larger sample of SNe~II. With this goal in mind, we
extend our previous work on SN II light curve modeling to a set of twenty
especially well-observed events. This allows us to measure the
mass and extent of the dense CSM, constrain
the zero-age main-sequence (ZAMS) masses, and measure the
explosion energies. From
this we can derive better constraints on just how common this dense CSM
is (at least $70\%$ of our sample), as well as look for correlations between
the various properties that may provide clues to the dense CSM's origin.
Beyond just the CSM properties, because we are able to constrain the
ZAMS masses for a wide sample, this provides important complementary
information about SN II progenitors to other studies of RSGs (for example,
work on pre-explosion imaging, \citealp{smartt:09,smartt:09a}).

In Section~\ref{setup}, we describe the details of our simulations. The sample of
twenty SNe II used for this work are presented in Section~\ref{numerical} along with our modeling
strategy and fitting results. In Section~\ref{discussion}, we explore what we can learn
with our full sample of fits as well as looking for correlations between different
properties of the SN progenitors that we measure. In Section~\ref{conclusions}, we summarize our
results and discuss future work.


\section{Numerical Setup}
\label{setup}

As in our previous work \citep{morozova:17} we use the non-rotating solar-metallicity 
RSG models from the stellar evolution code \texttt{KEPLER} 
\citep{weaver:78,woosley:07,woosley:15,sukhbold:14,sukhbold:16}. Above these
models we add a CSM extending out to a radius $R_{\rm ext}$ with the density
profile of a steady-state wind
\be
\label{wind}
	\rho(r) = \frac{\dot{M}}{4\pi r^2 v_{\rm wind}} = \frac{K}{r^2},
\ee
where $\dot{M}$ is the wind mass loss rate and $v_{\rm wind}$ is the wind
velocity. A steady-state wind is widely used in the literature to describe the structure 
of  CSM in the vicinity of RSGs \citep{chugai:07,
ofek:10,chevalier:11,moriya:11}, and we use it as a convenient prescription
to explore the diversity of possible CSM properties with just two parameters,
$K$ and $R_{\rm ext}$
(the case of an accelerating wind is considered in \citealp{moriya:17},
while for the cases of exponential or power-law density 
distributions see \citealp{nagy:16}).  We assume the
temperature and composition of the CSM to be constant and equal to their
values at the surface of the underlying RSG models. 
In this work, we do not address the physical mechanism responsible for the formation
of the CSM, which is crucial in defining the details of its structure. Instead, 
we concentrate on the general characteristics of the CSM, such as 
its total mass $M_{\rm CSM}$ and extent $R_{\rm ext}$. Numerical studies
of extended material around SNe demonstrate that the exact density distribution
only of secondary importance to these main properties \citep[e.g.,][]{piro:17}.

\begin{figure}
  \centering
  \includegraphics[width=0.475\textwidth]{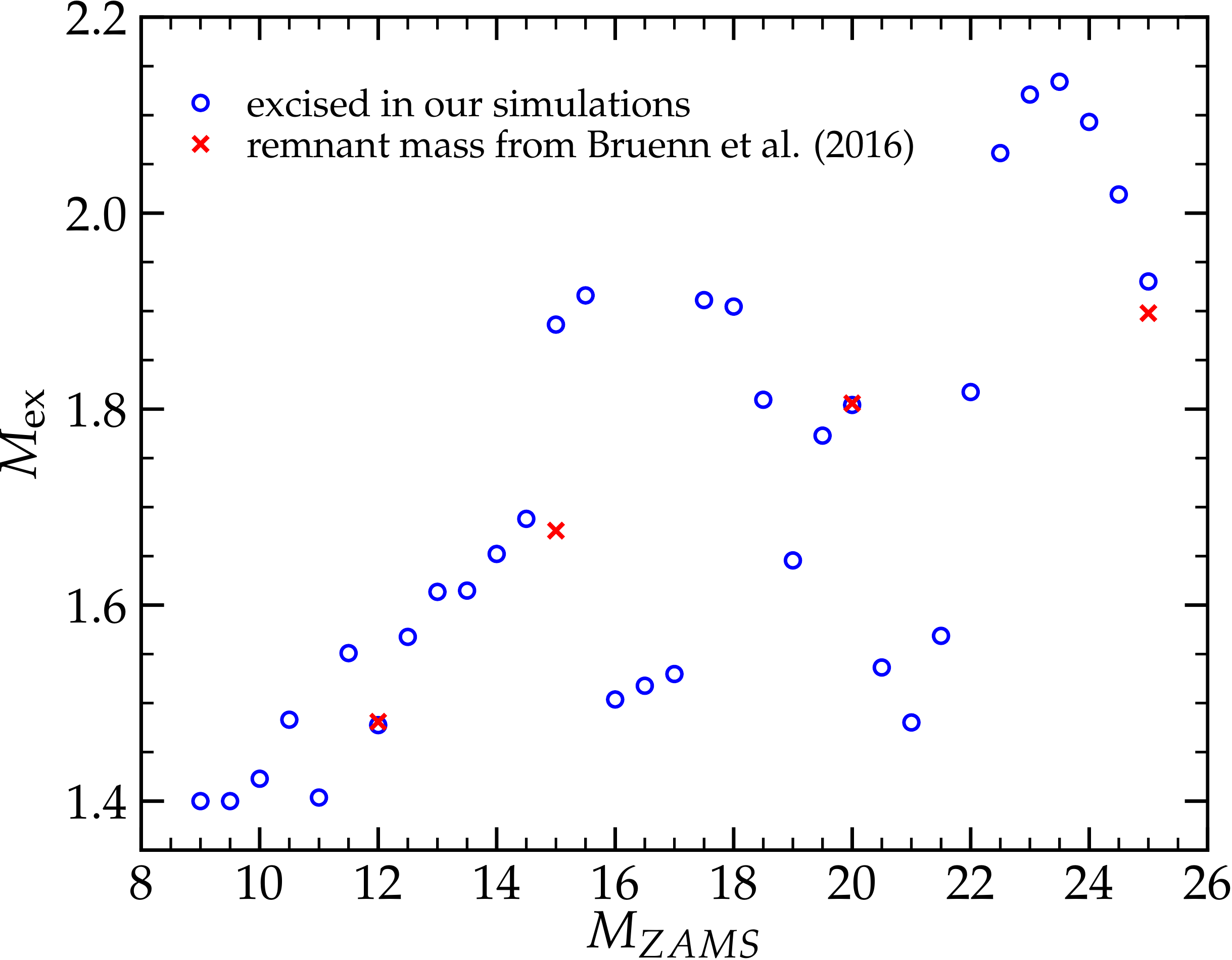}
  \caption{Excised mass as a function of ZAMS mass in our simulations 
  (blue circles), which is taken to be equal to the mass coordinate of the 
  silicon-oxygen interface in the pre-collapse composition profile. For
  comparison, in red we show the baryonic rest masses of 
  proto-neutron stars formed in core-collapse SN explosion mechanism simulations
  by \citet{bruenn:16} using progenitor models of \citet{woosley:07}.} 
  \label{fig:mex}
\end{figure}

From the large set of progenitor models presented in \citet{sukhbold:16},
we choose a subset of models in the mass range between $9\,M_{\odot}$
and $25\,M_{\odot}$ in steps of $0.5\,M_{\odot}$. Despite uncertainties in
the upper limit \citep{smartt:09a,smith:11,groh:13,dwarkadas:14}, the stars in this 
range of ZAMS masses are believed to be progenitors of the bulk of Type II SNe.
To obtain light curves from these models, we explode them with our open-source
numerical code \texttt{SNEC} \citep{morozova:15} in the range of 
asymptotic explosion energies, $E_{\rm fin}$, between $0.04$ and 
$1.3\,{\rm B}$ in steps of $0.02\,{\rm B}$, where 
$1\,{\rm B} = 10^{51}\,{\rm erg}$. We use a thermal bomb mechanism for the 
explosions, where the thermal bomb energy, $E_{\rm bomb}$ is found from $E_{\rm fin}$
and the total pre-explosion (negative, mostly gravitational) energy of the model, 
$E_{\rm init}$, as $E_{\rm bomb} = E_{\rm fin} - E_{\rm init}$. This energy is
then injected into the inner $0.02\,M_{\odot}$ of the model for a duration of
$1\,{\rm s}$. Previous work discusses the impact of this duration choice
\citep{morozova:15,morozova:16,morozova:17}. Exploding higher mass
progenitors with low energies is 
challenging for our code, since the current version is not capable of 
treating material falling back onto a remnant. Therefore, there is a region of 
the parameter space not covered in our study (shown in Figure~\ref{fig:chisq1} 
as a gray shaded area).

Before exploding the models, we excise the inner part, which is assumed to form
a neutron star. As shown in the large number of SN explosion mechanism 
simulations 
\citep[e.g.,][]{mueller:12,summa:16,suwa:16,burrows:16,radice:17}, 
the passage of the otherwise stalled 
shock wave through the density and composition discontinuities of the
progenitor model, such as the boundary between the silicon- and oxygen-
burning shells, can facilitate its revival and play an important role in the
successful SN explosion. Therefore, one can expect
the remnant masses to be equal or slightly larger than the mass coordinate of the Si/O 
interface in the progenitor composition profile, and we use this as a
criterium to determine the excised mass in our simulations. Figure~\ref{fig:mex}
shows the excised mass as a function of ZAMS 
mass for our subset of models. The values of 
the baryonic rest mass of the proto-neutron stars from the simulations of
\citet{bruenn:16}, using progenitor models from \citet{woosley:07}, 
are shown for comparison. Larger excision masses facilitate the explosion
of large ZAMS mass progenitors to some extent. We note thought that as
long as the explosion is successful,
the light curve depends very weakly on the excised mass.

In \texttt{SNEC}, we use the equation of state by \citet{paczynski:83} and
solve for the ionization fractions of hydrogen and helium following the
approach of \citet{zaghloul:00}. The numerical grid is identical to the one
used in our previous studies and consists of 1000 cells \citep{morozova:15,
morozova:16,morozova:17}. We use the same prescription for the opacity floor as
in those works, namely, $0.01\,{\rm cm}^2\,{\rm g}^{-1}$ for the solar metallicity
$Z=0.02$, $0.24\,{\rm cm}^2\,{\rm g}^{-1}$ for $Z=1$, with the linear 
dependence in between. We smoothen the composition profiles
before the explosion by passing a ``boxcar" with a width of 
$0.4\,M_{\odot}$ through the models four times. Photometric light curves
are calculated assuming black body emission and using the \texttt{MATLAB}
package for astronomy and astrophysics for
calculating specific wave bands \citep{ofek:14b}.


\section{Numerical models of 20 SNe~II}
\label{numerical}

\subsection{Observational data}
\label{data}

The SN sample we analyze here was chosen from 
a collection of high-quality SN~II light curves presented in
\citet{valenti:16}. The two main criteria we used in selecting
these were (i) good multi-band light curve
coverage and (ii) having an estimate of the 
$^{56}{\rm Ni}$ mass 
from the radioactive tail. We have omitted a few SNe due to
the large uncertainty in the explosion date ($>6$ days) as well
as the lack of data points during the transition between the
plateau and the radioactive tail, both of which are important to obtain 
good quality fits using our approach.

Table~\ref{tab:SN1} summarizes the observed properties of the
selected SN sample. The distance modulus, DM, together with 
its uncertainty, $\Delta$DM, are taken from Table 2 of \citet{valenti:16}.
The interstellar and galactic values
of the absorption in $B$-band, $A_{B, i}$ and $A_{B, g}$, respectively, are used to correct the
light curves for reddening according to the Cardelli law \citep{cardelli:89}
\footnote{Note that the interstellar reddening 
shown in Table~\ref{tab:SN1} was estimated based on the
equivalent width of Na \texttt{I} D lines in the spectra of host galaxies 
and does not take into account
possible extinction due to the circumstellar dust. However, since the
circumstellar dust is likely destroyed by the SN explosion, it should not
affect the light curve modeling. Negligible interstellar reddening for
some SNe from the sample is additionally supported by the fact that
they have similar colors during the plateau phase.}.
The plateau length $t_{\rm PT}$ is obtained by fitting a Fermi-Dirac
function to the transition between the plateau and radioactive tail, and
$\Delta t_{\rm PT}$ is its uncertainty\footnote{The values of $t_{\rm PT}$
are taken from Table D5 of \citet{valenti:16} when available, otherwise 
from Table D4.}. The $^{56}{\rm Ni}$ mass, $M_{\rm Ni}$, and its 
uncertainty, $\Delta M_{\rm Ni}$, were derived in \citet{valenti:16}
from the comparison of the post-plateau pseudo-bolometric light curves 
to the one of SN 1987A.

It has been highlighted in some observational work
that some SN light curves demonstrate a change in slope at a few
tens of days after the maximum \citep{anderson:14,valenti:16}.
This change is more pronounced in the
pseudo-bolometric light curves (including bands from $U$/$B$ to $I$) 
than in the single bands. The time $t_{\rm S}$
in Table~\ref{tab:SN1} marks the last data point used to measure the
early slope of the pseudo-bolometric light curves of the 
corresponding SNe in \citet{valenti:16} (the first parameter  
{\it ph\_stop} in their Table~D4), which we use here as a proxy
for the transition time between the early faster 
and the late shallow slopes.

\begin{table*}
\renewcommand{\arraystretch}{1.3}
\centering
\caption{Observed SN parameters. \label{tab:SN1}}
\begin{tabular}{lccccccccc}\hline \hline
SN & DM & $\Delta$DM & $A_{B, i}\,[{\rm mag}]$ & $A_{B, g}\,[{\rm mag}]$ 
& $t_{\rm PT}\,[{\rm d}]$  & $\Delta t_{\rm PT}\,[{\rm d}]$ & $t_{\rm S}\,[{\rm d}]$ 
& $M_{\rm Ni}\,[M_{\odot}]$& $\Delta M_{\rm Ni}\,[M_{\odot}]$\\
\hline
1999em & 30.34 & 0.07 & 0.234 & 0.174 & 118.1 & 1.0 & 34.4 & 0.0536 & 0.0119  \\ 

1999gi & 30.34 & 0.14 & 0.0 & 0.07 & 127.8 & 3.1 & 38.2 & 0.0320 & 0.0023 \\ 

2001X & 31.59 & 0.11 & 0.0 & 0.173 & 114.7 & 5.0 & 41.9 & 0.0550 & 0.0047 \\ 

2003Z & 31.70 & 0.15 & 0.0 & 0.141 & 124.2 & 4.5 & 29.4 & 0.0047 & 0.0002 \\ 

2003hn & 31.14 & 0.26 & 0.71 & 0.057 & 106.9 & 4.0 & 42.8 & 0.0324 & 0.0046 \\

2004et & 28.36 & 0.09 & 0.03 & 1.48 & 123.5 & 4.0 & 53.4 & 0.0414 & 0.0086 \\ 

2005cs & 29.26 & 0.33 & 0.0 & 0.205 & 126.0 & 0.5 & 22.3 & 0.0021 & 0.0002  \\ 

2009N & 31.67 & 0.11 & 0.532 & 0.078 & 108.3 & 1.2 & 26.6 & 0.0165 & 0.0021 \\ 

2009ib & 31.48 & 0.31 & 0.537 & 0.105 & 140.1 & 2.0 & 38.1 & 0.0520 & 0.0162  \\ 

2012A & 29.96 & 0.15 & 0.05 & 0.1 & 106.5 & 2.0 & 28.4 & 0.0087 & 0.0012 \\

2012aw & 29.96 & 0.09 & 0.24 & 0.115 & 135.2 & 4.0 & 36.6 & 0.0497 & 0.0059 \\ 

2012ec & 31.32 & 0.15 & 0.414 & 0.096 & 107.9 & 5.0 & 36.3 & 0.0394 & 0.0051 \\ 

2013ab & 31.90 & 0.08 & 0.081 & 0.099 & 101.8 & 1.0 & 36.4 & 0.0588 & 0.0100 \\ 

2013by & 30.81 & 0.15 & 0.0 & 0.798 & 85.4 & 2.0 & 39.7 & 0.0320 & 0.0043 \\ 

2013ej & 29.79 & 0.20 & 0.0 & 0.25 & 98.8 & 1.0 & 38.4 & 0.0207 & 0.0019 \\

LSQ13dpa & 35.08 & 0.15 & 0.0 & 0.137 & 128.7 & 2.0 & 36.9 & 0.0714 & 0.0127 \\ 

2014cy & 31.87 & 0.15 & 0.0 & 0.2 & 122.6 & 1.0 & 33.0 & 0.0037 & 0.0038 \\ 

ASASSN-14dq & 33.26 & 0.15 & 0.0 & 0.254 & 101.0 & 5.5 & 33.6 & 0.0461 & 0.0079 \\ 

ASASSN-14gm & 31.74 & 0.15 & 0.0 & 0.406 & 110.6 & 1.5 & 43.3 & 0.0767 & 0.0102 \\ 

ASASSN-14ha & 29.53 & 0.50 & 0.0 & 0.033 & 136.8 & 1.5 & 37.5 & 0.0014 & 0.0002 \\ 

\hline
\end{tabular}
\end{table*}

\subsection{Two step approach to numerical modeling}
\label{fitting}

As discussed in Section \ref{sec:introduction}, recent work
demonstrates that a dense, compact CSM is crucial for
 modeling Type II SN light curves \citep{morozova:17}.
This means that to properly fit these light curves requires fitting
for $M_{\rm ZAMS}$, $E_{\rm fin}$, $M_{\rm Ni}$, $^{56}$Ni
mixing, explosion time, $R_{\rm ext}$, and $K$--a seven parameter fitting space!
Since building such a larger grid of models is simply
not feasible, we take advantage of the knowledge that the
early light curve over the first $\sim10-30\,{\rm days}$ should be
dominated by the CSM, and that the remainder of the light curve is
dominated by the hydrogen-rich RSG envelope. This allows
us to utilize a two
step approach in fitting the SN light curves, which we describe
further below.

In the first step of our light curve fitting, we generate
a grid of light curves in $M_{\rm ZAMS}-E_{\rm fin}$ parameter
space, in each case using the radioactive $^{56}{\rm Ni}$
mass for each SN from Table~\ref{tab:SN1}. 
As was discussed in Section~\ref{setup}, we cover the 
parameter space $9\,M_{\odot}<M_{\rm ZAMS}<25\,M_{\odot}$
in steps of $0.5\,M_{\odot}$ and 
$0.04\,{\rm B}<E_{\rm fin}<1.3\,{\rm B}$ in steps of $0.02\,{\rm B}$
(with the exception of the largest masses and smallest energies).
To account for the effect of $^{56}{\rm Ni}$ mixing
into the envelope, we consider three degrees of $^{56}{\rm Ni}$ mixing for
each SN, up to the mass coordinates of $3\,M_{\odot}$, $5\,M_{\odot}$
and $7\,M_{\odot}$ (in each case using a ``boxcar'' method as described above).

Within this grid, we look for the best fitting model for each SN by
minimizing $\chi^2$, which we calculate as
\begin{equation}
\label{chisq}
\chi^2 = \sum_{\lambda\in[g,...,z]} \,\,\, 
\sum_{t_{\rm S}<t^*<t_{\rm PT}}\frac{(M_{\lambda}^*(t^*)-M_{\lambda}(t^*))^2}{(\Delta M_{\lambda}^*(t^*))^2}\ ,
\end{equation}
where $M_{\lambda}^*(t^*)$ is the observed magnitude in
a given band $\lambda$ at the moment of observation $t^*$,
$\Delta M_{\lambda}^*(t^*)$ is the 
corresponding observational error, and
$M_{\lambda}(t^*)$ is the numerically obtained magnitude
in the same band at the same moment of time. 
To take into account uncertainties in the explosion date, we shift the time
of observations within the allowed range of explosion times in steps of 
$0.5\,{\rm d}$, and look for the minimal $\chi^2$.
In the first fitting
step, we use only the parts of light curves between the 
time $t_{\rm S}$ and the end of plateau $t_{\rm PT}$.
We do not go beyond $t_{\rm PT}$, since after this time the
whole ejecta is expected to become optically thin and the
diffusion approach to radiation transport used in \texttt{SNEC}
is not valid anymore. We also do not include $u$-, $U$- and
$B$- bands in Equation~\ref{chisq}, since after day $\sim 20$ 
the radiation in these bands is affected by iron group line blanketing \citep{kasen:09},
which is not taken into account. As a result of this first step, we get the best fit values of 
progenitor ZAMS mass, explosion energy, degree of $^{56}{\rm Ni}$ mixing
and explosion time.

For the second step, we attach CSM on top of these
best fitting RSG models, 
we generate a grid of light curves 
in $R_{\rm ext}-K$ parameter space for each SN. We vary $K$ in the
range between $1.0\times10^{17}$ and $3.0\times10^{18}$ in steps
of $1.0\times10^{17}$, and $R_{\rm ext}$ in the range between $700\,R_{\odot}$
and $3800\,R_{\odot}$ in steps of $100\,R_{\odot}$. We shift the
observational data with respect to the explosion date in the same way that
minimized $\chi^2$ during the first step. After that we assess the best 
fitting model within the $R_{\rm ext}-K$ grid by calculating $\chi^2$ as we did
in Equation~\ref{chisq}, but for $t^*<t_{\rm PT}$ instead of $t_{\rm S}<t^*<t_{\rm PT}$.
This second step results in the best fitting values of the CSM
density parameter $K$ and external radius $R_{\rm ext}$.

\subsection{Numerical results}
\label{results}

\begin{figure*}
  \centering
  \includegraphics[width=0.9\textwidth]{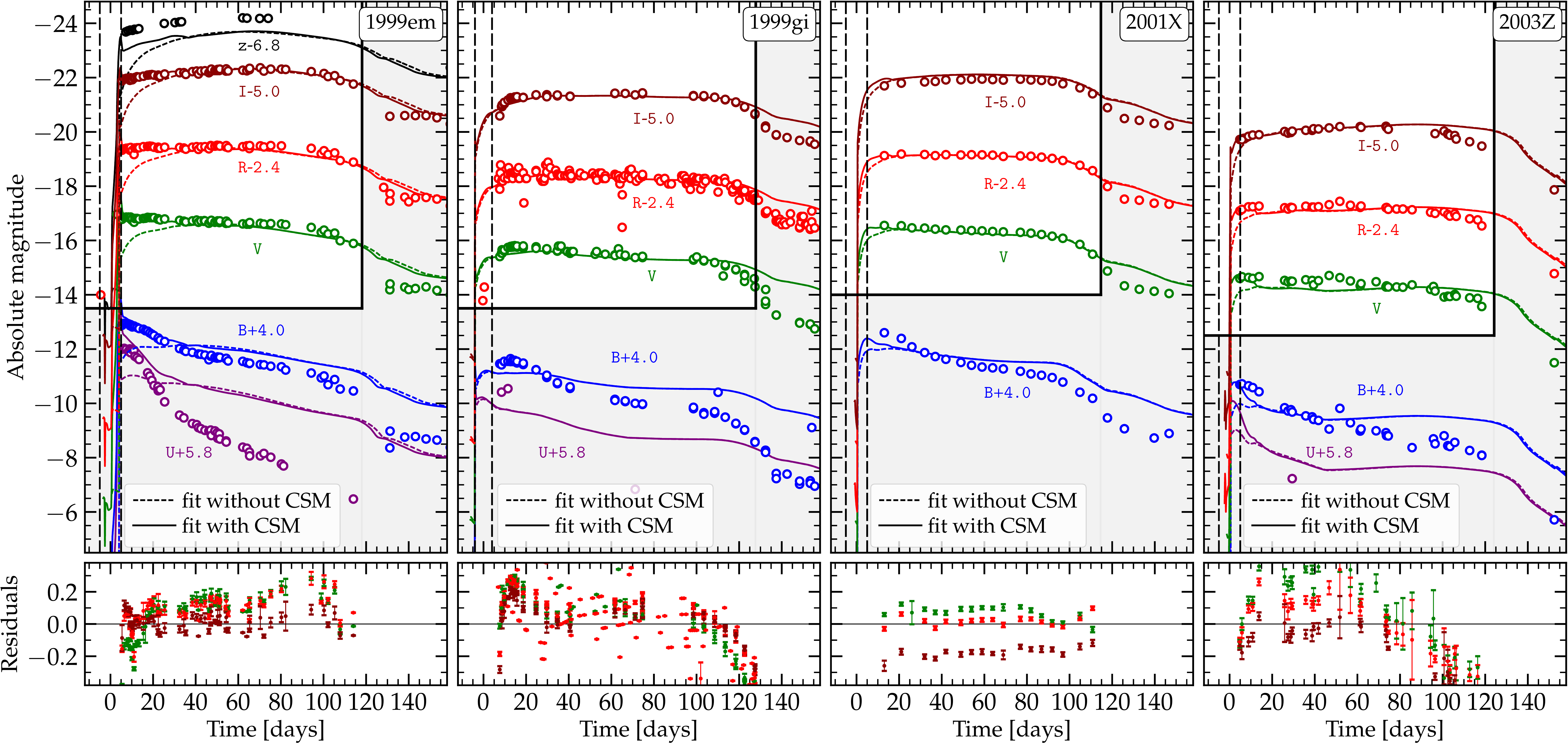}
  \includegraphics[width=0.9\textwidth]{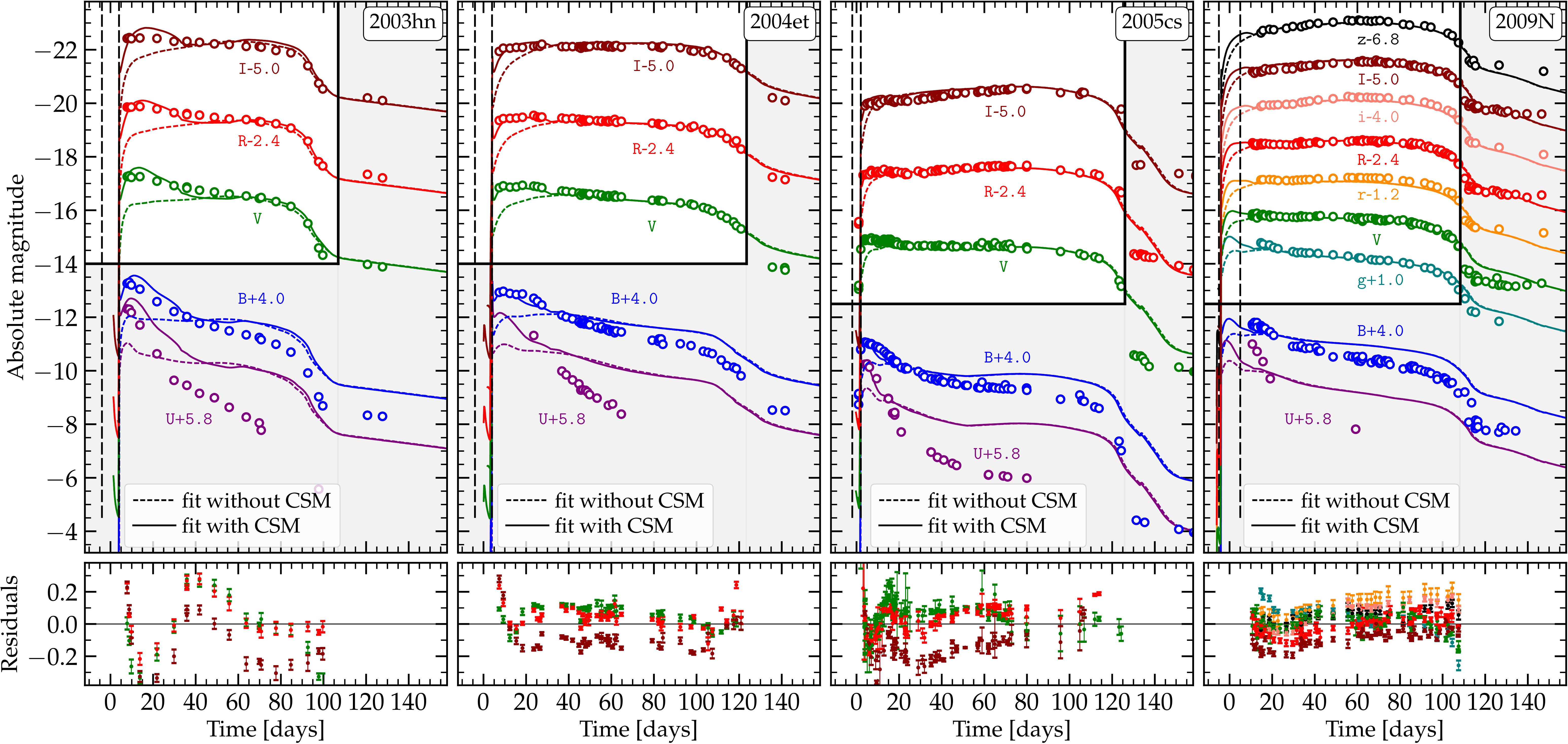}
  \includegraphics[width=0.9\textwidth]{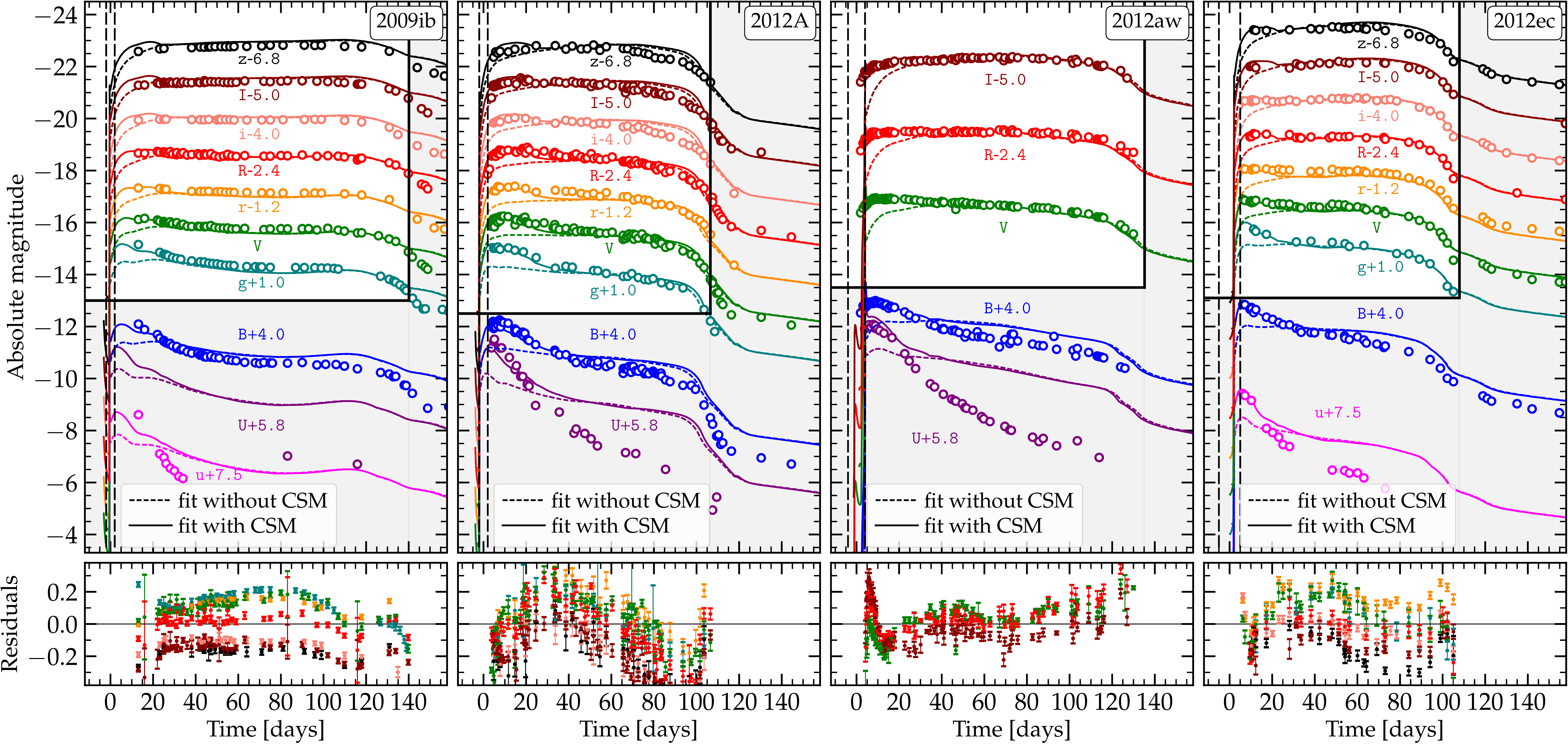}
  \caption{Best fit light curves with (solid lines) and without (dashed lines)
  CSM. Data is shown as open circles. Each color corresponds to a different
  wave band as labeled. Unshaded (white) regions contain the SN data used to find the fits
  while shaded (gray) regions are ignored for fitting either because it is too late in the light curve
  or iron group line blanketing may be important (in the bluer) bands.} 
  \label{fig:fits1}
\end{figure*}
\begin{figure*}
  \centering
  \includegraphics[width=0.9\textwidth]{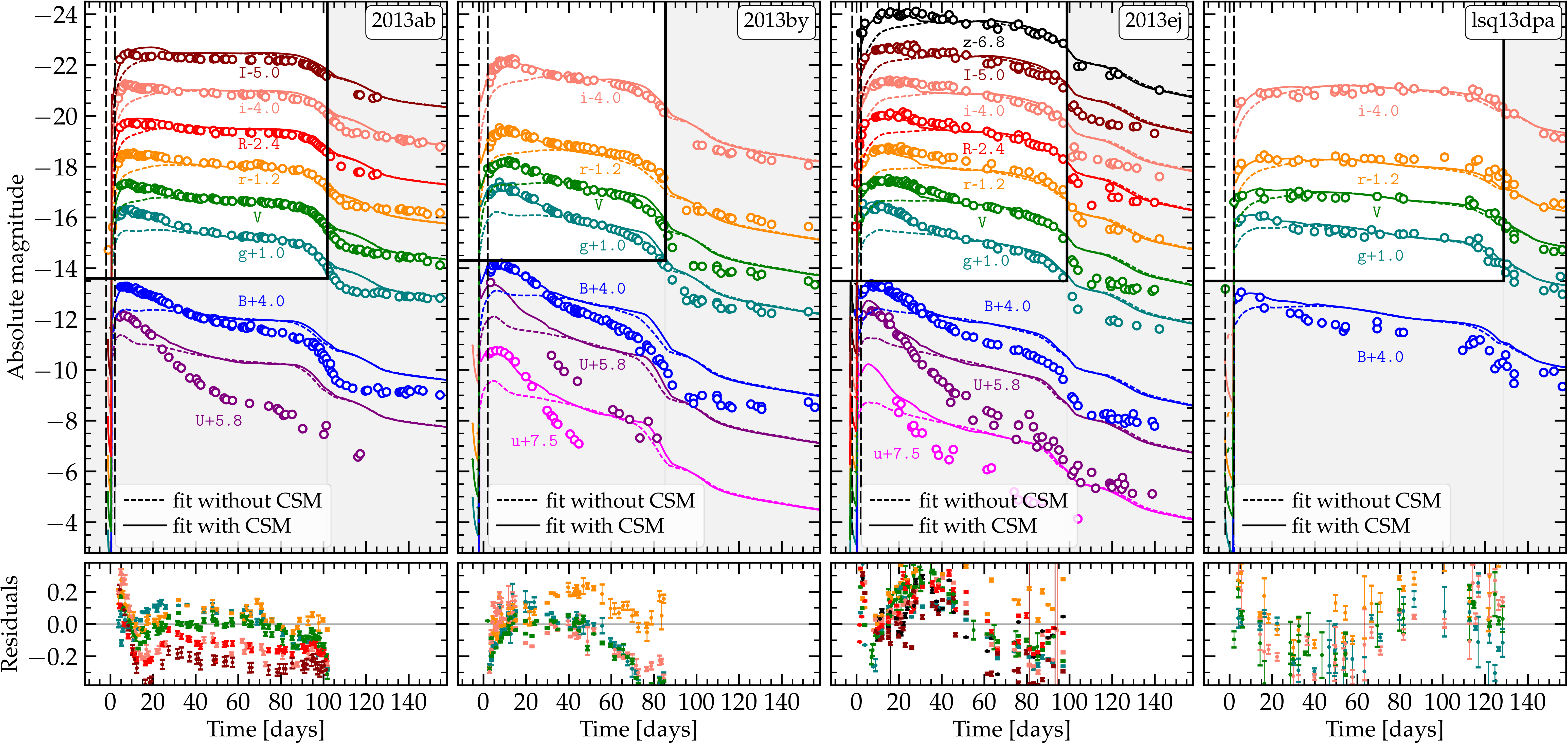}
  \includegraphics[width=0.9\textwidth]{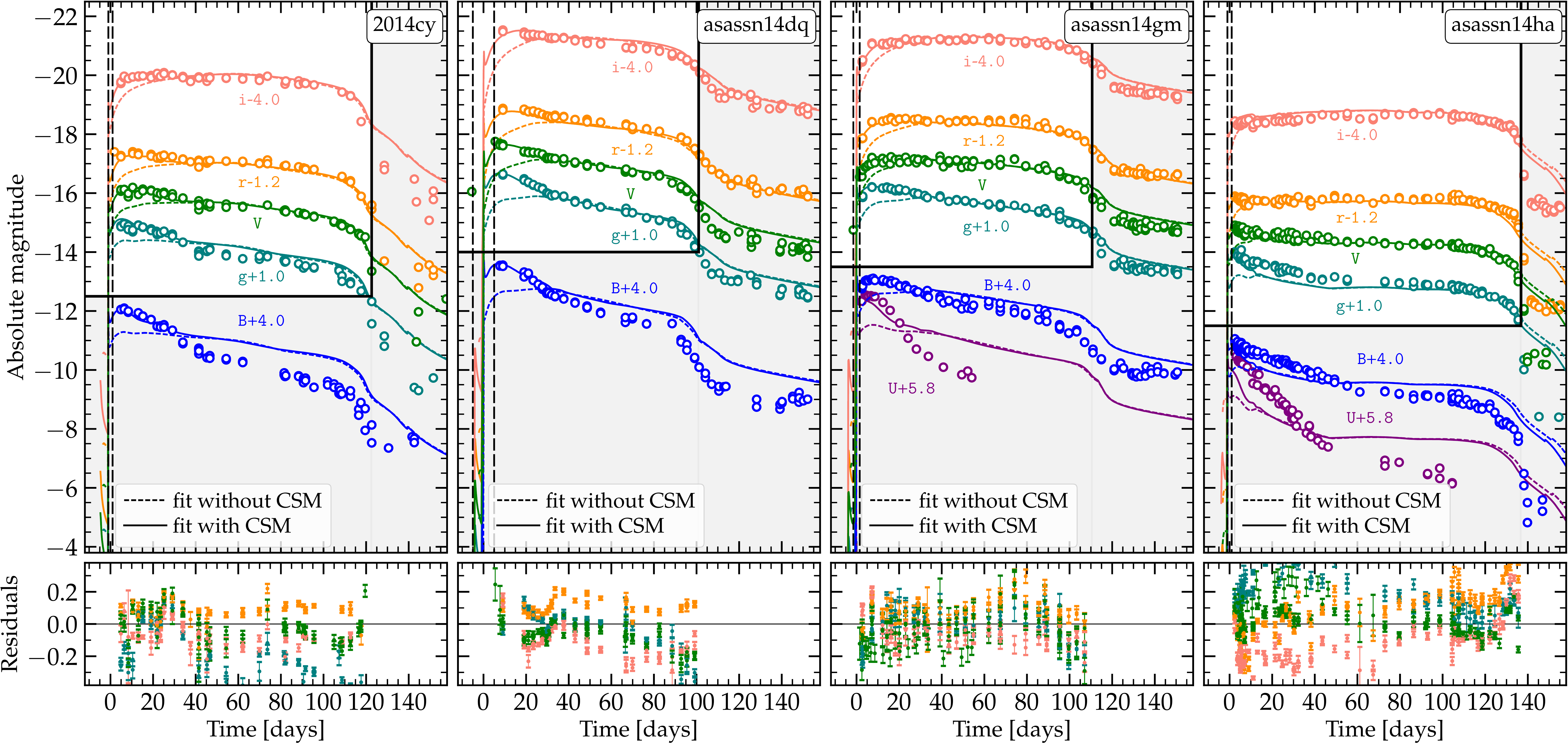}
  \caption{Additional SNe, the same as Figure \ref{fig:fits1}.} 
  \label{fig:fits2}
\end{figure*}

Figures~\ref{fig:fits1} and~\ref{fig:fits2} show all of the best fitting
models for the SNe from our set, together with the observed 
light curves in different bands. The dashed lines show the result of 
the first fitting step (without CSM), while the solid lines show
the final fit (with CSM). Residuals are shown to assess the quality
of the fits. The shaded regions in the plots contain the data that were not used 
in our analysis for the reasons described in Section~\ref{fitting}.
For completeness, we partially show the radioactive tails of our 
light curves. We emphasize, however, that our code uses diffusion approach
for the radiation transport, which works well for the shock cooling
and plateau part of the light curve, but is not suitable for the
nebular phase. In addition, as described in the code manual 
\footnote{\url{https://stellarcollapse.org/codes/snec_notes-1.00.pdf}}, for the purpose of
magnitude calculations the effective temperature is kept
above $5000\,{\rm K}$, following the reasoning of \citet{swartz:91}.
Therefore, no conclusion
concerning the goodness of the fit can be made based
on data after the end of plateau.

%
\begin{figure*}
  \centering
  \includegraphics[width=0.95\textwidth]{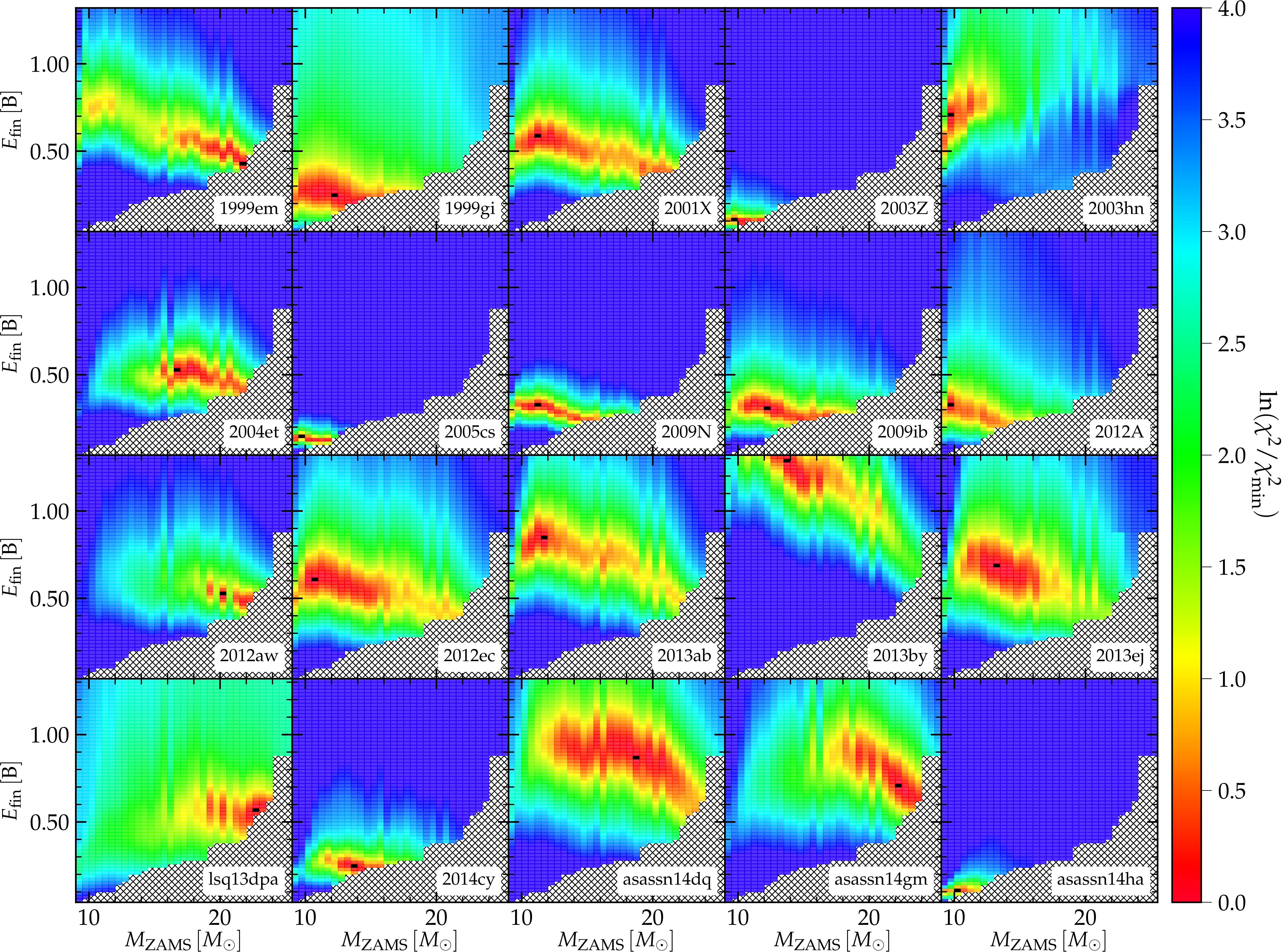}
  \caption{Color coded ratio of $\chi^2$ to $\chi^2_{\rm min}$ at each
  grid point in $M_{\rm ZAMS}-E_{\rm fin}$ space. Yellow contour
  approximately corresponds to the $\chi^2$ increase by a factor of 
  $e$ with respect to the minimum.The black square in each panel 
  indicates the best fitting parameters. Gray shaded regions 
  could not be covered by the current study due to the numerical
  difficulties. For each SN, the plotted $\chi^2$ corresponds to the
  best fitting degree of $^{56}{\rm Ni}$.} 
  \label{fig:chisq1}
\end{figure*}
\begin{figure*}
  \centering
  \includegraphics[width=0.95\textwidth]{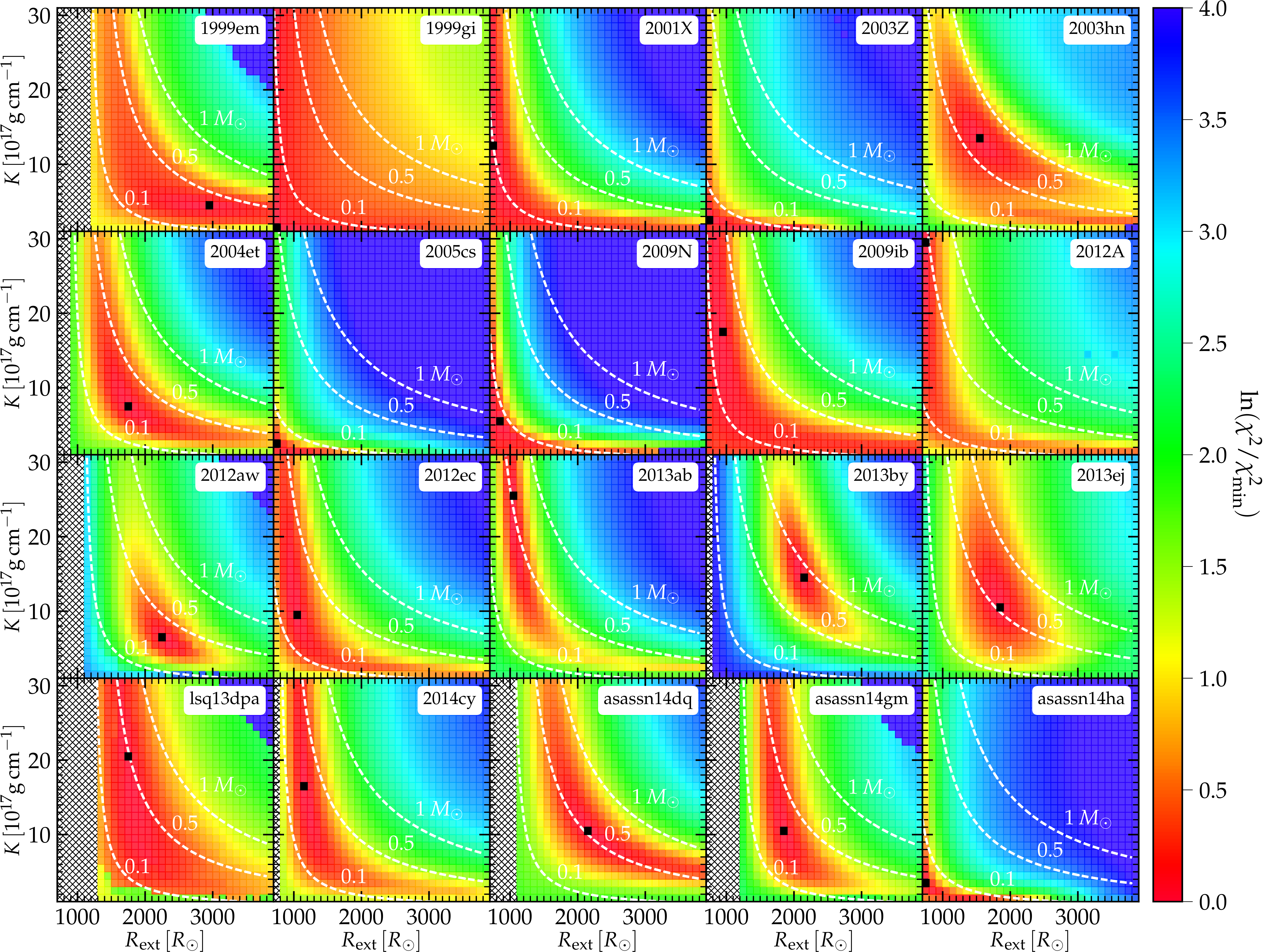}
  \caption{Color coded ratio of $\chi^2$ to $\chi^2_{\rm min}$ at each
  grid point in $R_{\rm ext}-K$ space. Yellow contour
  approximately corresponds to the $\chi^2$ increase by a factor of 
  $e$ with respect to the minimum.The black square in each panel 
  indicates the best fitting parameters. Shaded with gray are the regions 
  below the RSG radius of the best fitting model for the corresponding
  SNe. White lines show the constant CSM mass contours 
  ($0.1\,M_{\odot}$, $0.5\,M_{\odot}$ and $1\,M_{\odot}$).} 
  \label{fig:chisq2}
\end{figure*}

As emphasized in previous work \citep{morozova:17}, we find that light
curves obtained from the bare RSGs are significantly poorer at
fitting the data in comparison to light curves including dense CSM.
The dashed curves in Figures~\ref{fig:fits1} and~\ref{fig:fits2} rise more slowly
than the observations during the first $20-40$ days (similar to Figures~2-4 of 
\citealt{morozova:17}). On the other hand, if one were to restrict the fitting window for these
models to just the time between the slope break and the end of plateau, then
reasonable fits can be made with just bare RSGs. This lends support for the two step
method we utilized for the light curve fitting.
Interestingly, the addition of
dense CSM even improves the fit of the early light curve in the $B$ and $U$
bands that we do not include when fitting.
This is not surprising given that
at the high effective temperatures of the early light
curves ($10,000\,{\rm K}$ and higher) the iron group line blanketing does
not play an important role and the spectrum is very close to a black body.
Nevertheless, this is further evidence for the necessary presence of dense CSM.

Figures~\ref{fig:chisq1} and~\ref{fig:chisq2} illustrate the dependence of $\chi^2$ 
on the grid parameters. Since the minimum of $\chi^2$ in our models is rather 
shallow, it does not make much sense to access the confidence regions as we previously
did in \citet{morozova:17}, because the $39.3\%$ confidence region will cover
$\sim39.3\%$ of the domain, etc. Instead, we color code the natural
logarithm of $\chi^2/\chi^2_{\rm min}$, so that the yellow contour in the figures
approximately corresponds to the increase of $\chi^2$ by a factor of $e$ with respect
to the minimum, while the green region corresponds to an order of magnitude 
increase. We adopt the extents of the yellow contours as the error estimates of
our fits. Shaded gray regions in Figure~\ref{fig:chisq1} indicate the part of the
domain which could not be covered by the light curves due to numerical 
difficulties, while in Figure~\ref{fig:chisq2} they just cut away the regions smaller
than the radii of the underlying RSGs.

From Figure~\ref{fig:chisq1} one can see that the explosion energy is generally
better constrained by our models than the progenitor ZAMS mass. The 
pre-explosion profiles of the progenitors are determined by
complex nuclear burning processes during stellar evolution and do not always demonstrate
smooth dependence on the ZAMS mass, which results in striped patterns
in Figure~\ref{fig:chisq1}. Figure~\ref{fig:chisq2} demonstrates a strong
degeneracy between the density and the radial extent of the CSM. It is
interesting to note, however, that the contours of constant $\chi^2$
follow very closely the contours of constant CSM mass, which we plotted
in the figure with white dashed lines. This tells us that the total mass of the
CSM can be inferred from our fits more robustly than its density and
external radius separately (as has been found for
extended material in other types of SNe, \citealp{piro:17}). As for the CSM radius,
it is only a few times the RSG radius in all cases. This indicates that whatever causes the mass loss
must happen at the very last stages of the RSG's life.

All fitting parameters are summarized in Table~\ref{tab:SN2}. 
The different degrees of $^{56}{\rm Ni}$
mixing result in rather close, and in some cases even identical, fitting progenitor
characteristics (the best fitting model among the three mixing degrees
is shown in bold). The energy released in the radioactive decay of $^{56}{\rm Ni}$
is expected to prolong the plateau and/or flatten it \citep{nakar:16}. 
In our grid, though, this subtle effect gets smeared by others, like
discontinuities in the properties of the progenitor models themselves, and the
difference in fitting parameters between the different mixing degrees is smaller than the 
error bars we estimate based on Figure~\ref{fig:chisq1}.

As for the CSM, we added to the table its total mass $M_{\rm CSM}$,
without trying to convert it into the mass losses in $M_{\odot}\,{\rm yr}^{-1}$,
which an interested reader can easily estimate from Equation~(\ref{wind})
for a reasonable wind speed. In fact, there is no reason
to assume a wind speed of $10\,{\rm km}\,{\rm s}^{-1}$, which is commonly
observed in steady state RSG winds. Higher wind velocities of 
$\sim$100$\,{\rm km}\,{\rm s}^{-1}$ in our models would correspond to larger mass loss rates 
of the order of few $M_{\odot}\,{\rm yr}^{-1}$ and the duration of the enhanced 
mass loss period of only few months before explosion. However, since our
models are not sensitive to the wind velocity in the pre-explosion models, but
rather to the density profile of the wind, we cannot put any constraint on the
mass loss rate and the duration of the outflow. Moreover, we emphasize that
it is entirely possible that the dense CSM is not exactly a wind but a different
density distribution for which the wind profile we use is an approximation. The origin
of the CSM is an interesting topic of investigation on its own, and it lays
beyond the scope of this study, although we discuss it further below.

\begin{table*}
\renewcommand{\arraystretch}{1.3}
\centering
\caption{Best fit parameters. Bold parameters highlight those with the
smallest $\chi^2$ among the three degrees of $^{56}{\rm Ni}$ mixing.
These values are shown in Figures~\ref{fig:fits1}-\ref{fig:chisq2} of
Section~\ref{numerical} and used for the analysis of 
Section~\ref{discussion}. All masses and radii are given in solar units,
energies in ${\rm B}$ and parameter $K$ in ${\rm g}\,{\rm cm}^{-1}$. The last column
lists the values used in Figure~\ref{fig:summary} only and not meant to 
fully represent the numerous literature (see more references in the text). \label{tab:SN2}}
\begin{tabular}{l|cc|cc|cc|ccc|cc}\hline \hline
\multicolumn{1}{c|} {} & \multicolumn{2}{c|}{$^{56}{\rm Ni}$ mixed to $3\,M_{\odot}$} 
& \multicolumn{2}{c|}{$^{56}{\rm Ni}$ mixed to $5\,M_{\odot}$}
& \multicolumn{2}{c|}{$^{56}{\rm Ni}$ mixed to $7\,M_{\odot}$}&
\multicolumn{3}{c|}{CSM parameters} & \multicolumn{2}{c}{Best fit $M_{\rm ej}$}\\
SN &  $M_{\rm ZAMS}$&
$E_{\rm fin}$& $M_{\rm ZAMS}$ &
$E_{\rm fin}$ & $M_{\rm ZAMS}$ &
$E_{\rm fin}$ & $K$ & $R_{\rm ext}$ &
$M_{\rm CSM}$ & This work & Other works \\
\hline
1999em & {\bf 21.5} & {\bf 0.42} & 20.5 & 0.52 & 20.0 & 0.48 & $4.0\times 10^{17}$ & 2900 & 0.31 & 14.48 & 19\textcolor{blue}{$^{a}$} \\ 

1999gi & {\bf 12} & {\bf 0.24} & 12 & 0.22 & 10.5 & 0.22 & $1.0\times 10^{17}$ & 700 & $<0.003$ & 9.42 & - \\ 

2001X & {\bf 11} & {\bf 0.58} & 12 & 0.5 & 12.0 & 0.46 & $1.2\times 10^{18}$ & 700 & $<0.07$ & 9.29 & - \\ 

2003Z & {\bf 9.5} & {\bf 0.1} & 10 & 0.08 & 10 & 0.08 & $2.0\times 10^{17}$ & 700 & $<0.03$ & 7.81 & 14\textcolor{blue}{$^{a}$}, 11.3\textcolor{blue}{$^{b}$} \\ 

2003hn & 9.5 & 0.76 & {\bf 9.5} & {\bf 0.7} & 9.5 & 0.66 & $1.3\times 10^{18}$ & 1500 & 0.63 & 7.81 & - \\

2004et & 20 & 0.42 & 18 & 0.54 & {\bf 16.5} & {\bf 0.52} & $7.0\times 10^{17}$ & 1700 & 0.25 & 12.47 & 22.9\textcolor{blue}{$^{a}$} \\ 

2005cs & 9.5 & 0.14 & 9.5 & 0.14 & {\bf 9.5} & {\bf 0.14} & $2.0\times 10^{17}$ & 700 & $<0.03$ & 7.81 & 15.9\textcolor{blue}{$^{a}$}, 9.5\textcolor{blue}{$^{c}$} \\ 

2009N & 10 & 0.34 & 11 & 0.34 & {\bf 11} & {\bf 0.32} & $5.0\times 10^{17}$ & 800 & 0.05 & 9.29 & 11.5\textcolor{blue}{$^{d}$} \\ 

2009ib & {\bf 12} & {\bf 0.3} & 14.5 & 0.24 & 12.5 & 0.24 & $1.7\times 10^{18}$ & 900 & 0.2 & 9.42 & 15\textcolor{blue}{$^{e}$} \\ 

2012A & {\bf 9.5} & {\bf 0.32} & 9.5 & 0.3 & 9.5 & 0.3 & $2.9\times 10^{18}$ & 700 & $<0.38$ & 7.81 & 12.5\textcolor{blue}{$^{f}$} \\

2012aw & 20 & 0.52 & 20 & 0.52 & {\bf 20} & {\bf 0.52} & $6.0\times 10^{17}$ & 2200 & 0.3 & 14.04 & 19.6\textcolor{blue}{$^{g}$} \\ 

2012ec & 10.5 & 0.68 & 10.5 & 0.64 & {\bf 10.5} & {\bf 0.6} & $9.0\times 10^{17}$ & 1000 & 0.18 & 8.71 & - \\ 

2013ab & {\bf 11.5} & {\bf 0.84} & 12 & 0.76 & 11.5 & 0.7 & $2.5\times 10^{18}$ & 1000 & 0.48 & 9.20 & 7\textcolor{blue}{$^{h}$} \\ 

2013by & {\bf 13.5} & {\bf 1.28} & 12.5 & 1.3 & 12.5 & 1.28 & $1.4\times 10^{18}$ & 2100 & 0.83 & 10.16 & -  \\ 

2013ej & {\bf 13} & {\bf 0.68} & 13.5 & 0.66 & 13 & 0.66 & $1.0\times 10^{18}$ & 1800 & 0.49 & 9.95 & 10.6\textcolor{blue}{$^{i}$} \\

LSQ13dpa & 20 & 0.56 & 20 & 0.56 & {\bf 22.5} & {\bf 0.56} & $2.0\times 10^{18}$ & 1700 & 0.43  & 14.30 & - \\ 

2014cy & 13.5 & 0.24 & 13.5 & 0.24 & {\bf 13.5} & {\bf 0.24} & $1.6\times 10^{18}$ & 1100 & 0.25 & 10.16 & - \\ 

ASASSN-14dq & 19.5 & 0.86 & {\bf 18.5} & {\bf 0.86} & 18.5 & 0.86 & $1.0\times 10^{18}$ & 2100 & 0.48 & 13.13 & - \\ 

ASASSN-14gm & 23.5 & 0.62 & 23.5 & 0.62 & {\bf 22} & {\bf 0.7} & $1.0\times 10^{18}$ & 1800 & 0.27 & 14.40 & - \\ 

ASASSN-14ha & {\bf 10} & {\bf 0.1} & 10 & 0.1 & 10 & 0.1 & $3.0\times 10^{17}$ & 700 & $<0.03$ & 8.25 & - \\ 

\hline
\end{tabular}
    \\ 
    $^{a}$\citet{utrobin:13}, $^{b}$\citet{pumo:17}, $^{c}$\citet{spiro:14}, $^{d}$\citet{takats:14}, $^{e}$\citet{takats:15},\\ $^{f}$\citet{tomasella:13}, $^{g}$\citet{dallora:14}, $^{h}$\citet{bose:15}, $^{i}$\citet{huang:15}
\end{table*}

It is worth noting that in  \citet{morozova:17} we also modeled two SNe from the current sample, 2013ej and
2013by, but using a slightly
different approach. There we generated entire 4-dimensional grids of light
curves in $M_{\rm ZAMS}$, $E_{\rm fin}$, $K$, $R_{\rm ext}$
parameter space, instead
of using the two-step fitting procedure described here. This previous approach 
was more computationally expensive, which translated to a coarser
resolution of these parameters.
In \citet{morozova:17}, for SN 2013ej we got the
values of $M_{\rm ZAMS}=12.5\,M_{\odot}$, $E_{\rm fin}=0.6\,{\rm B}$,
$K=1.0\times 10^{18}\,{\rm g}\,{\rm cm}^{-1}$, $R_{\rm ext}=2100\,R_{\odot}$, and for SN 2013by we got
$M_{\rm ZAMS}=14.5\,M_{\odot}$, $E_{\rm fin}=1.4\,{\rm B}$,
$K=1.0\times 10^{18}\,{\rm g}\,{\rm cm}^{-1}$, $R_{\rm ext}=2300\,R_{\odot}$, both of
which are in good agreement with the
values from Table~\ref{tab:SN2}. 
This adds additional credence to the scheme we use in the
current work. Unfortunately, we could not fit SN 2013fs using the two-step approach because
of the lack of data near the transition between the
plateau and radioactive tail.


\section{Discussion}
\label{discussion}

With the results from Section \ref{results}, we have for the first
time ZAMS masses and dense CSM properties for a collection of
Type II SNe based solely on light curve fitting. This allows us to
compare the parameters we measure with other studies in the
literature using different methods to constrain the properties of
SN II progenitors. Furthermore, we can
look for correlations between the various parameters with the
hope of getting a better understanding of the explosion mechanism
as well as the origin of the dense CSM. We discuss each of
these comparisons below.

\subsection{Ejecta masses}

\begin{figure}
  \centering
  \includegraphics[width=0.475\textwidth]{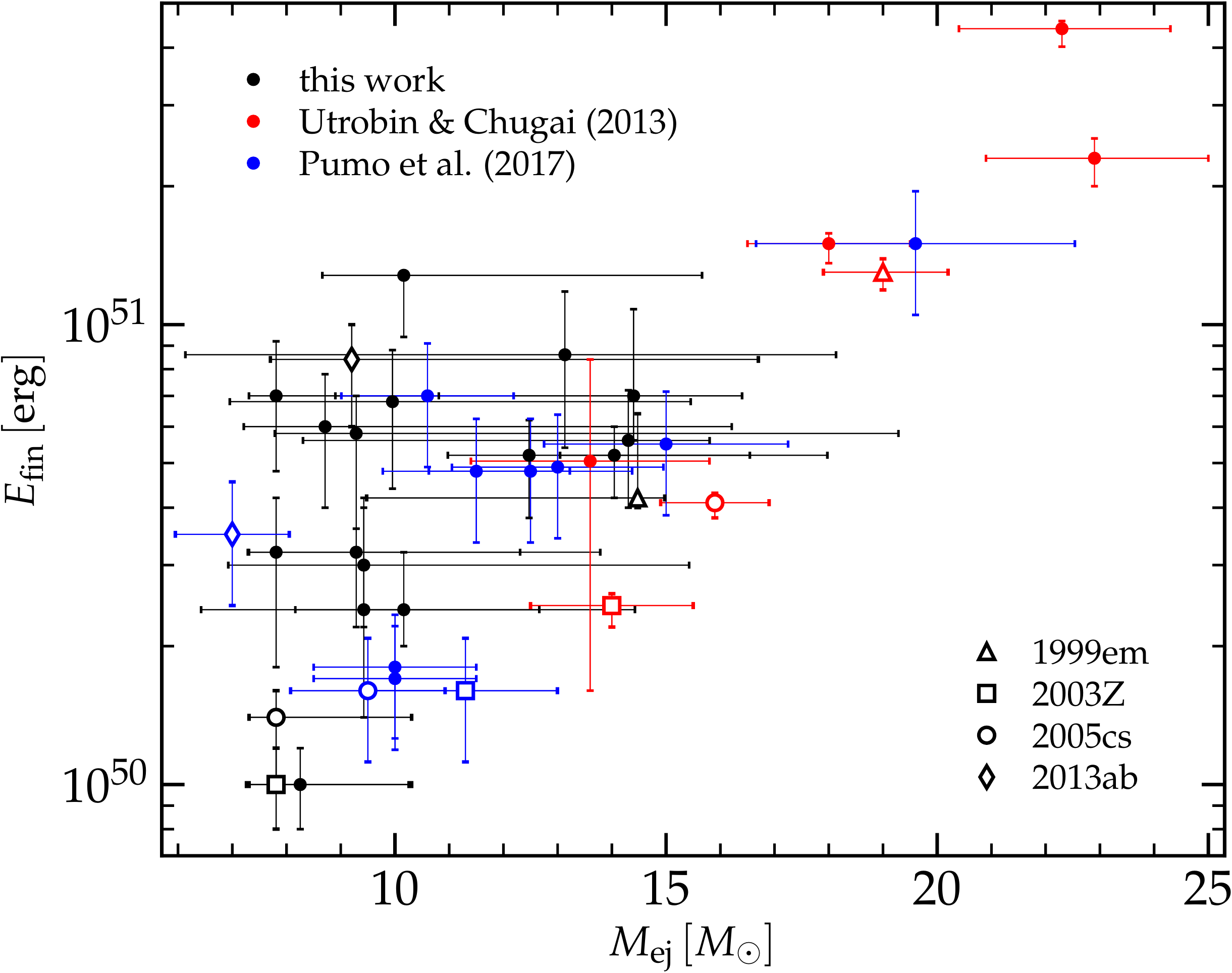}
  \caption{Ejecta masses versus explosion energies derived for our 
   sample of 20 SNe (black symbols), compared to the results of previous works.
  The error bars for our values are estimated based on the extent of the
  yellow contours in Figure~\ref{fig:chisq1}, for \citet{utrobin:13}
  based on their Figure~8, and for \citet{pumo:17} taken to 
  be $30\%$ of the value of energy and $15\%$ of the value of mass, as suggested
  by the authors.} 
  \label{fig:summary}
\end{figure}

Figure~\ref{fig:summary} compares the ejecta masses and explosion 
energies obtained in our work to the other recently published studies
of \citet{utrobin:13} and \citet{pumo:17} (who, together with their own, 
collect the results of \citealt{tomasella:13,spiro:14,dallora:14,takats:14,
takats:15,bose:15,huang:15}). This figure contains only the values obtained
from the hydrodynamical models of the corresponding events, and
does not include the numerous estimates obtained from the analytical scalings, spectra
or X-ray/radio signals \citep[see, among many others,][]{misra:07,jerkstrand:12,
jerkstrand:14,chakraborti:16,yuan:16,dhungana:16}. 
These values are collected in the last column of Table~\ref{tab:SN2}. 
Some of the SNe that
are common between samples are marked with special symbols.

\begin{figure}
  \centering
  \includegraphics[width=0.465\textwidth]{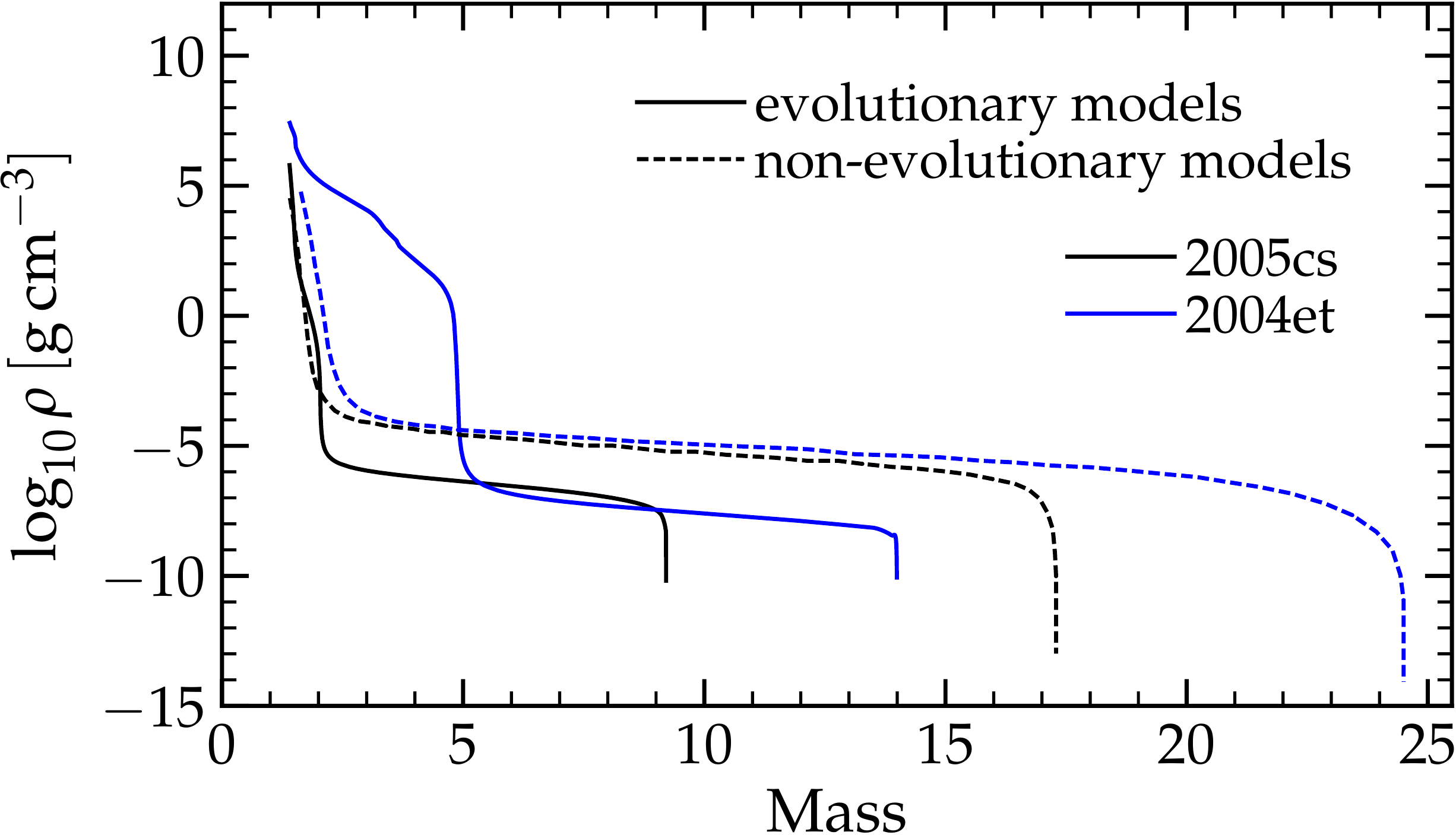}
  \caption{Comparison of the evolutionary and non-evolutionary models
  that produce numerical light curves  (using different
  explosion energies) that compare favorably to the observations
  of SNe 2005cs and 2004et.} 
  \label{fig:profiles}
\end{figure}

For almost all of the SNe (with the exception of SN 2013ab) we get lower
ejecta masses and energies than the previous work. 
At least part of this difference
comes from us using stellar evolution progenitor models instead of the double
polytropic models widely used in the literature. The difference 
between evolutionary and non-evolutionary models that are capable of fitting the
observational data equally well has been
investigated by \citet{utrobin:17} for SN 1999em. We add to this
by comparing non-evolutionary and evolutionary models for SNe
2005cs and 2004et in Figure~\ref{fig:profiles}. For the evolutionary profiles
we show our best fit models from Section~\ref{numerical}, while the
polytropic models are from \citet{utrobin:08} and \citet{utrobin:09}.
It is seen from Figure~\ref{fig:profiles} that the
non-evolutionary models predict considerably higher progenitor masses
than the evolutionary models. It is known from the analytical scalings
\citep{arnett:80,chugai:91,popov:93}, that there is a certain degeneracy
in the way the ejecta mass and the explosion energy influence the
luminosity and duration of the light curve. Increasing the explosion energy
makes the plateau shorter but more luminous, and so does decreasing the
ejecta mass. This explains why more massive non-evolutionary
progenitors also require higher explosion energies 
in order to reproduce the observed light curves.
It is worth noticing that the difference in the fitting models shown 
in Figure~\ref{fig:profiles} is not likely due to the difference in the numerical
codes, because with \texttt{SNEC} we could successfully reproduce the 
fit of SN 1999em from \citet{bersten:11} (very similar to the one 
from \citealt{utrobin:07}), using the same double
polytropic model (see Appendix A of \citealt{morozova:15}).

Our results from Figure~\ref{fig:summary} show less of a correlation between
ejecta mass and explosion energy, but at the same time,
all three samples show a lack of high ejecta masses
with low energies. Although it is true that we did not consider some
regions of this parameter space due to the
numerical issues, from Figure~\ref{fig:chisq1} it is seen that most
of the best fitting models are not near this boundary of the
modeled region. Earlier, correlation between
the progenitor mass and the explosion energy was studied, for example, in
\citet{poznanski:13}.


\subsection{Progenitor ZAMS masses}
\label{zams}

It is interesting to examine our results in view of the RSG problem,
first described in \citet{smartt:09a}. The problem is that  analysis
of the archival pre-explosion images of SNe~IIP implies an
upper limit of $\sim17\,M_{\odot}$ for the ZAMS mass of their progenitors.
This is significantly lower than the expected limit of $\sim25\,M_{\odot}$, 
obtained from the comparison of the observed properties of RSGs with their
possible evolutionary tracks \citep{levesque:05,levesque:06}. In addition,
models of light curves of the same SNe are known to predict 
higher progenitor masses than those derived from the direct imaging
\citep{utrobin:08,utrobin:09}.

The reason for this discrepancy may lie either on the observational side, on
the theoretical side, or both. On the observational side, \citet{walmswell:12} proposed that extinction
due to the circumstellar dust helps to increase the progenitor masses
derived from the pre-explosion images. However, in the subsequent work, 
\citet{kochanek:15} demonstrated that the effects of dust composition, photon
scattering and near-IR dust emission may instead lead to an even lower progenitor
mass estimate. On the theoretical side, the problem may come from the
progenitor models or various physical approximations used in the codes.
Finally, it is possible that stars with ZAMS masses in the range $17-25\,M_{\odot}$ may evolve
past RSG stage due to enhanced mass loss or binary interactions and
eventually explode as SNe of other types rather normal Type II
(IIb/n or even Ib/c; see \citealt{smith:11}).

The upper limit for the ZAMS mass of SNe II progenitors could signify that more massive
stars collapse to black holes. They might eject their hydrogen envelope
in a low energy explosion (as described in the work of \citealp{nadezhin:80,piro:13,lovegrove:13,lovegrove:17}
and potentially observed by \citealp{adams:17}) rather than a typical SN.
Such a process would then imply that the helium core masses of these stars become the black
hole masses \citep{kochanek:14,clausen:15}. Therefore the inferred
maximum mass for SN II progenitors  may or may not explain
the observed distribution of Galactic black hole masses \citep{ozel:10} and the origin
of a potential mass gap between neutron stars and black holes \citep{ozel:12}.

\begin{figure}
  \centering
  \includegraphics[width=0.46\textwidth]{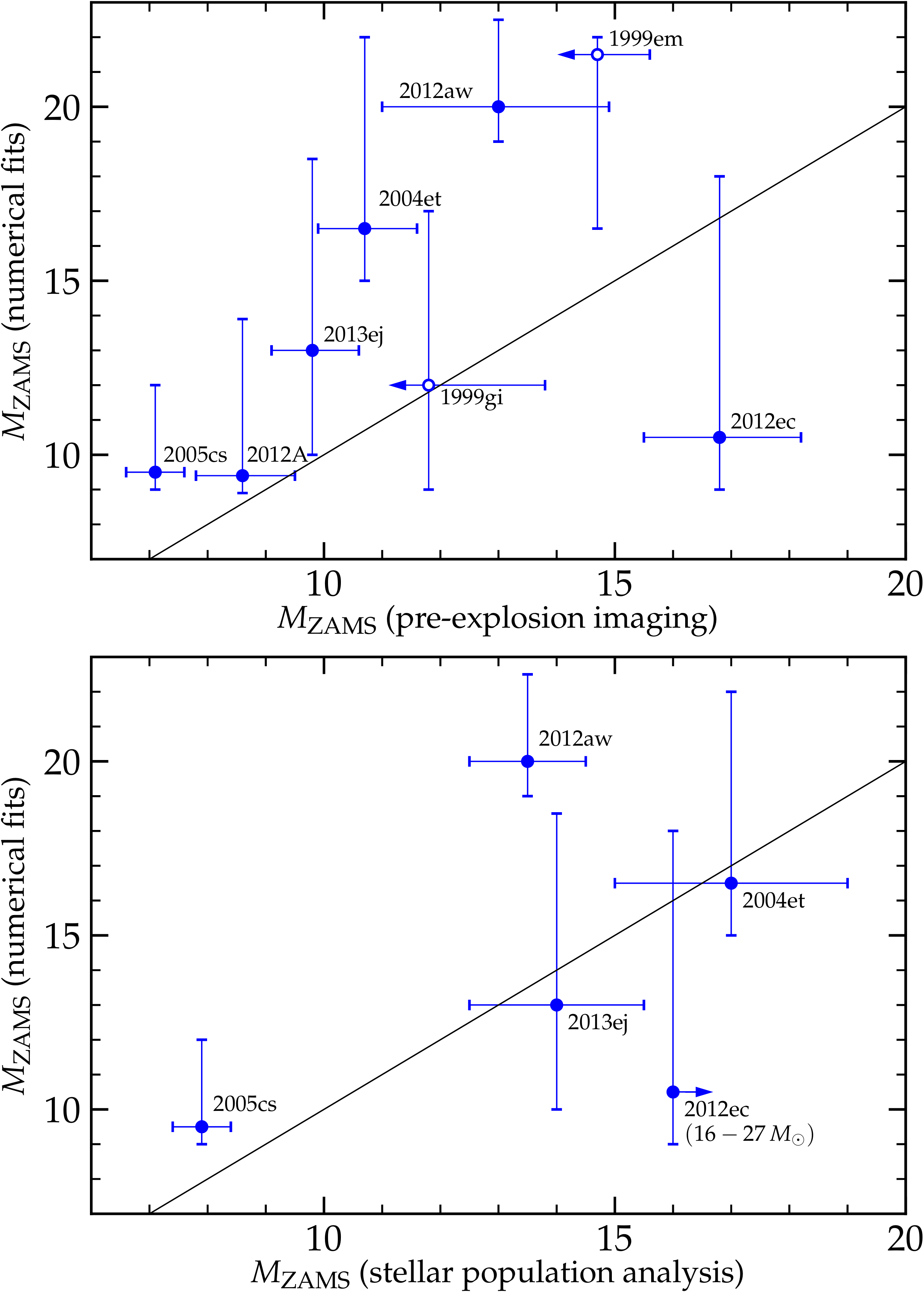}
  \caption{Top panel: Comparison of ZAMS masses of 8 SNe obtained from our numerical fits
  with those obtained with the pre-explosion imaging (\citealp{davies:18}, filled symbols indicate the
  progenitor detections, while empty symbols with arrows indicate upper limits from progenitor 
  non-detections). Poor agreement
  between the ZAMS masses may be at least partly explained by the difference between
  the progenitor models from \texttt{STARS} and \texttt{KEPLER} evolutionary codes 
  (\citealp{jerkstrand:14}, see the discussion in the text). Bottom panel: Comparison of ZAMS 
  masses of SNe obtained from our numerical fits with those obtained from the analysis of the surrounding
  stellar populations by \citet{maund:17}.} 
  \label{fig:preexp}
\end{figure}

The top panel of Figure~\ref{fig:preexp} shows 8 SNe from our sample
that also have an estimate of the progenitor ZAMS mass obtained from the
pre-explosion imaging. All values shown along the $x$-axis of 
Figure~\ref{fig:preexp} are taken from the recent work of \citet{davies:18}, which summarizes
and corrects the previous pre-explosion imaging mass estimates using
updated bolometric corrections for RSGs. Earlier analysis of the pre-explosion
images for the SNe shown in Figure~\ref{fig:preexp} may be found in 
\citet{walmswell:12,tomasella:13,maund:13,maund:14a,fraser:14,fraser:16,kochanek:15} 
and \citet{mauerhan:17}. 
We use filled symbols for SNe with direct progenitor detections
and the empty symbols for SNe that only have upper limits due to
a non-detection of the progenitor. 

At a first glance, the agreement between the masses
obtained from the pre-explosion imaging and the hydrodynamical modeling 
in Figure~\ref{fig:preexp} does
not look good. However, it is important to note that these two parts of the analysis 
currently use progenitor models from different stellar evolution codes (\texttt{STARS} is
used in \citet{davies:18}, and \texttt{KEPLER} in our work). The progenitor
models from these two evolutionary codes are compared in Figure~5 of \citet{jerkstrand:14}, which
demonstrates that for the same pre-explosion luminosity the \texttt{STARS} models
predict systematically lower ZAMS masses. The difference between the 
\texttt{STARS} and \texttt{KEPLER} masses
is about $1-2\,M_{\odot}$ in the lower ZAMS mass range ($10-15\,M_{\odot}$) and up to 
$4\,M_{\odot}$ in the higher ZAMS mass range ($20-25\,M_{\odot}$). Were the ZAMS masses in
Figure~\ref{fig:preexp} shifted correspondingly to the right along the $x$-axis, this would improve
the agreement with the results of our modeling. We note that our
$M_{\rm ZAMS}$ values for most of the SNe are lower than some of the other
ones previously obtained in the literature \citep{maguire:10,dallora:14}.

The bottom panel of Figure~\ref{fig:preexp} shows 5 SNe from 
our sample that have an estimate of the progenitor ZAMS mass obtained from the
analysis of the surrounding stellar populations by \citet{maund:17}. In this panel, SNe 2013ej and
2004et show better agreement with the numerical models than in the top panel. 
An interesting exception is SN 2012ec, for which 
the analysis of both pre-explosion images and stellar populations 
predicts considerably higher ZAMS mass than the numerical modeling.



%

\begin{figure}
  \centering
  \includegraphics[width=0.465\textwidth]{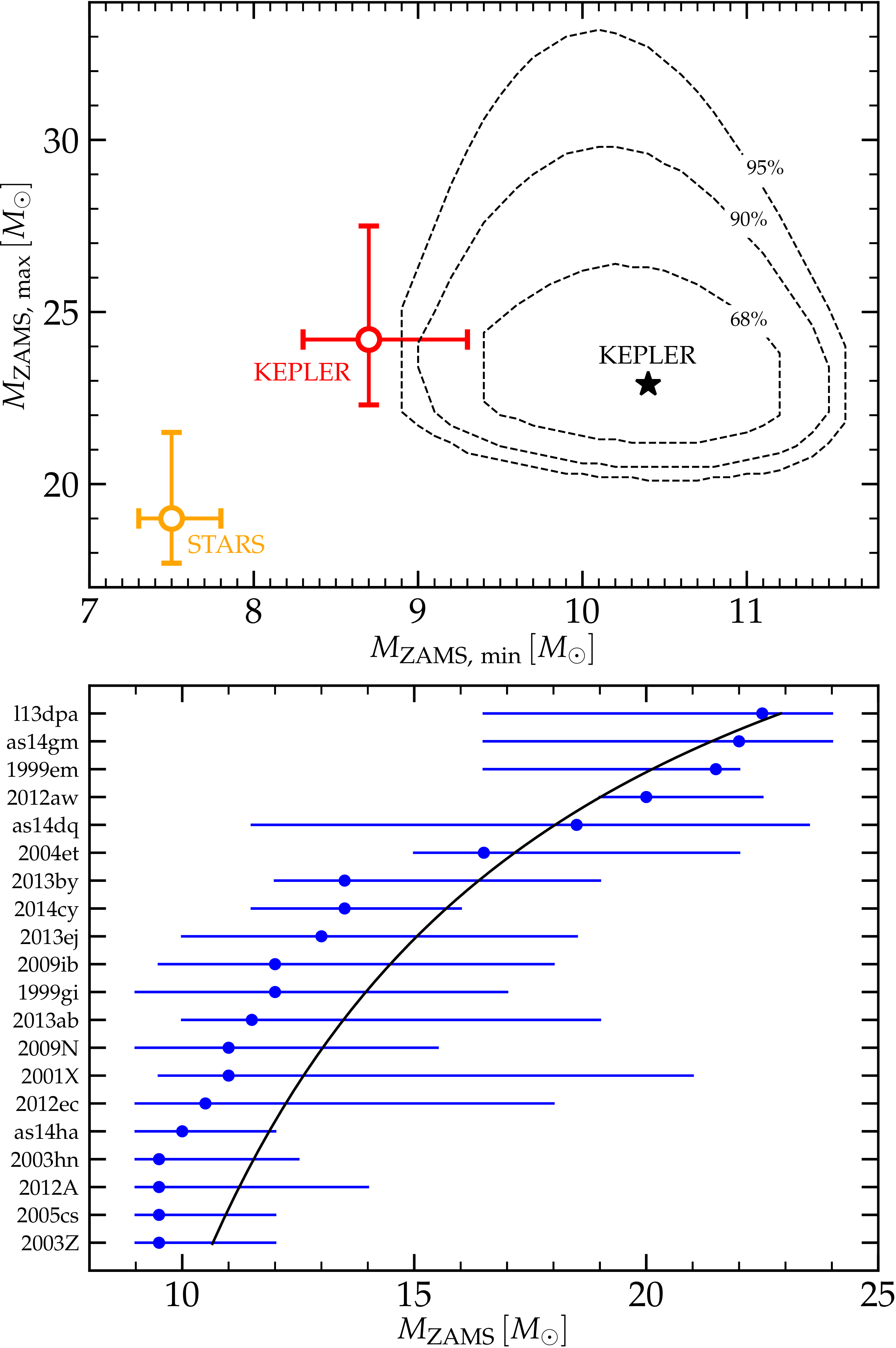}
  \caption{Top panel: The likelihood function for the minimum and
  maximum ZAMS masses of the Type~II SN progenitors. 
  The maximum likelihood derived from our numerical fits corresponds
  to $M_{\rm ZAMS,\ min}=10.4\,M_{\odot}$ and 
  $M_{\rm ZAMS,\ max}=22.9\,M_{\odot}$, indicated by a black star in the
  plot. The contours show the $68$, $90$ and $95\%$ confidence regions.
  For comparison, in yellow and red we show the values obtained in \citet{davies:18}
  from the pre-explosion imaging,
  using \texttt{STARS} and \texttt{KEPLER} stellar evolution codes, respectively. The 
  estimates obtained with the same evolutionary code demonstrate better
  agreement, supporting our earlier discussion on Figure~\ref{fig:preexp}.
  Bottom panel: The cumulative frequency plot of the ZAMS masses
  derived from the numerical fits. The black line shows a Salpeter IMF
  with minimum and maximum ZAMS masses of $10.4$ and 
  $22.9\,M_{\odot}$, respectively.} 
  \label{fig:likelihood}
\end{figure}

We next explore what ZAMS mass distribution is implied by our sample.
To do this we consider a Salpeter initial mass function (IMF) and investigate
what lower and upper limits on the progenitor ZAMS mass, $M_{\rm min}$ and 
$M_{\rm max}$, respectively, are needed to match our distribution.
Following \citet{smartt:09a},
we calculate the probability function for each SN in Table~\ref{tab:SN2} as
\begin{eqnarray}
P_j = && \int_{M_{j,\rm low}}^{M_j}\frac{\left(M-M_{j,\rm low}\right)M_j^{\Gamma-1}}
{\left(M_j-M_{j,\rm low}\right)\left(M_{\rm min}^{\Gamma}-M_{\rm max}^{\Gamma}\right)}
dM \nonumber \\ &&
+\int_{M_j}^{M_{j,\rm high}}\frac{M^{\Gamma-1}}
{\left(M_{\rm min}^{\Gamma}-M_{\rm max}^{\Gamma}\right)}dM \ ,
\end{eqnarray}
where $M_j$ is the best fitting ZAMS mass of the $j$-th SN,
$M_{j,\rm low}$ and $M_{j,\rm high}$ are its lower and upper uncertainties,
and $\Gamma=-1.35$ for the Salpeter initial mass function (IMF). When
$M_{j,\rm low}$ or $M_{j,\rm high}$ are higher or lower than $M_{\rm min}$ 
or $M_{\rm max}$, we use the latter as the integration limits. The 
maximum of the likelihood, calculated as $\mathcal{L} = \prod P_j$,
corresponds to $M_{\rm min}=10.4\,M_{\odot}$ and 
$M_{\rm max}=22.9\,M_{\odot}$, as shown in black in the top panel of 
Figure~\ref{fig:likelihood}. The $68$, $90$ and $95\%$ confidence regions
are estimated from the condition
\begin{equation}
\ln \mathcal{L}_{\rm max}-\ln \mathcal{L} = \frac{1}{2}\chi \ ,
\end{equation}
where $\chi=2.3$, $4.6$ and $6.2$, correspondingly. 
The bottom panel of Figure~\ref{fig:likelihood} shows the cumulative
frequency plot of the ZAMS masses derived from the numerical fit,
analogous to the plot first published in \citet{smartt:09a} for the ZAMS
masses derived from the pre-explosion images (their Figure 8). The solid
line shows the Salpeter IMF with the minimum mass of $10.4\,M_{\odot}$ 
and the maximum mass of $22.9\,M_{\odot}$, derived from maximizing
the likelihood. 

For comparison with the pre-explosion imaging, 
in the top panel of Figure~\ref{fig:likelihood} we show the values for
maximum and minimum masses obtained in \citet{davies:18}, using 
\texttt{STARS} and \texttt{KEPLER} stellar evolution codes. This plot
demonstrates that using the same set of progenitor models for the
pre-explosion imaging and the hydrodynamical modeling improves the
agreement between the mass limits.
The maximum mass that we obtain is closer to the maximum mass of
RSGs ($25-30\,M_{\odot}$) seen from observations \citep{massey:01,
levesque:05,levesque:06,crowther:07}. It therefore appears, coming back to the
discussion about RSGs above, that our distribution of masses relieves some of the
tension in the RSG problem.


%

\subsection{SN explosion properties}

\begin{figure}
  \centering
  \includegraphics[width=0.475\textwidth]{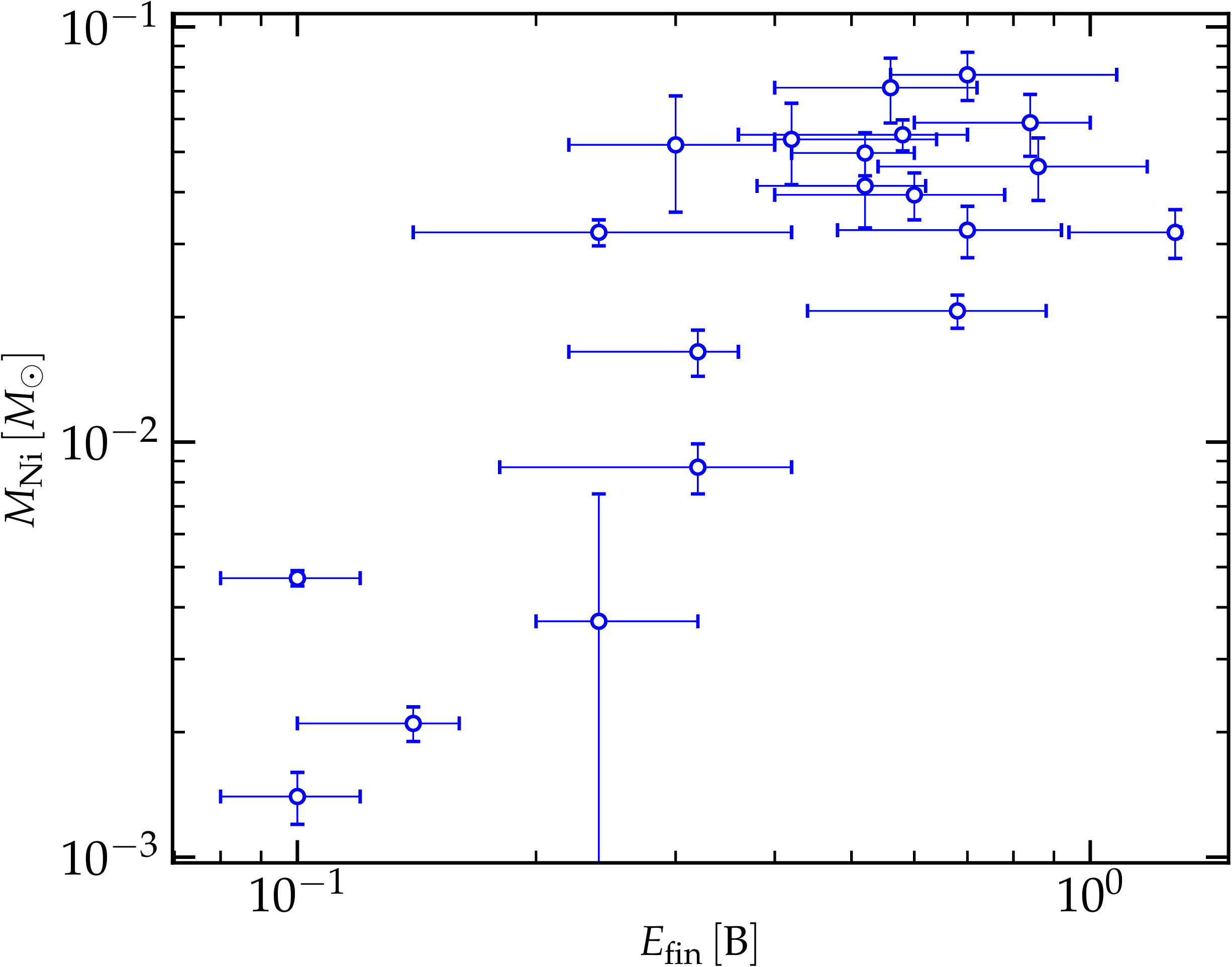}
  \caption{Correlation between the mass of $^{56}{\rm Ni}$ and 
  the explosion energy, obtained from our numerical fits.} 
  \label{fig:MNi}
\end{figure}

It is natural to assume that there must be a 
correlation between the explosion energy and
the $^{56}{\rm Ni}$ mass since this is synthesized
as the shock propagates out through the dense regions
of the core \citep{nadyozhin:03}.
This correlation has already been
seen by the nucleosynthetic simulations 
\citep{sukhbold:16} and light curve parameterization methods \citep{pejcha:15b}.
Furthermore, observations show this correlations through the 
$^{56}{\rm Ni}$ mass and the plateau magnitudes of SNe~II
\citep{hamuy:03,spiro:14,valenti:16} as well as their ejecta velocities
\citep{maguire:12}. Here we test whether this correlation is also
seen simply from our light curve fitting.
Figure~\ref{fig:MNi} compares the explosion
energy derived from the numerical fits and the $^{56}{\rm Ni}$ mass of
the corresponding SNe. This shows a strong correlation between the
two, similar to these other previous studies.

\subsection{CSM properties}

With our sample of SNe and models we can next address whether
the CSM mass and radii we infer correlate with other properties of the SN
or progenitor. This can hopefully provide clues about the mechanism by
which the dense CSM we are finding is generated.

\begin{figure}
  \centering
  \includegraphics[width=0.475\textwidth]{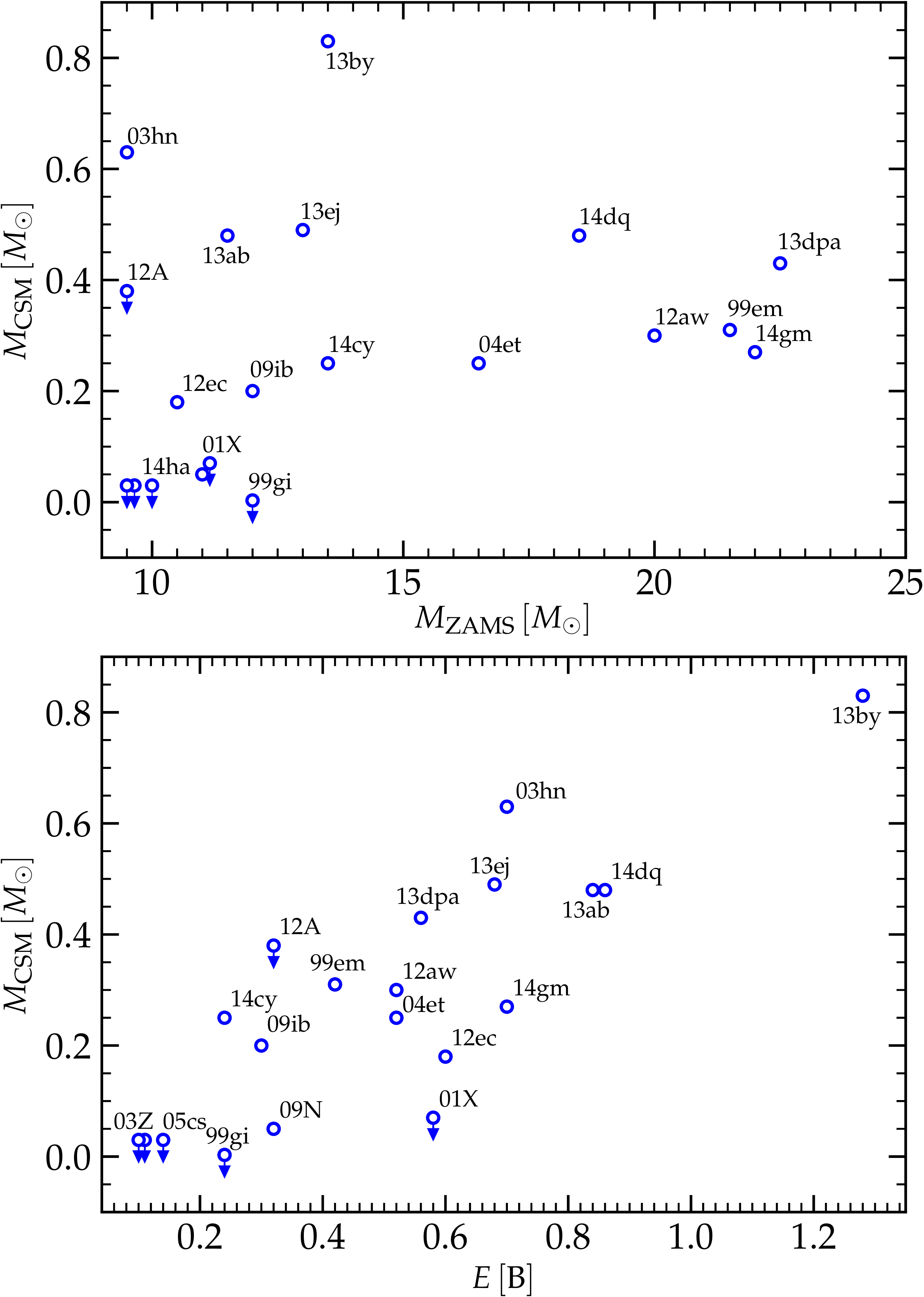}
  \caption{CSM mass versus ZAMS mass (top panel) and
  explosion energy (bottom panel), obtained from our 
  numerical light curve fits. Arrows indicate SNe for which we only
  have upper limits on CSM mass from Table~\ref{tab:SN2}.} 
  \label{fig:CSM}
\end{figure}

Figure~\ref{fig:CSM} shows the derived CSM masses versus
the ZAMS masses of the progenitors and the explosion energies.
The arrows indicate events that only have an upper
limit on the CSM mass. From the top panel, there does not appear
to be a strong correlation between $M_{\rm CSM}$ and $M_{\rm ZAMS}$.
Nevertheless, all of the events where we only found CSM mass upper limits
are below $M_{\rm ZAMS}\sim12\,M_\odot$, which may be a result of
our modeling or due to something physical that should be further explored.

On the other hand, the lower panel of Figure~\ref{fig:CSM}
suggests that there is indeed a correlation between $M_{\rm CSM}$
and the explosion energy. This general correlation was to be expected,
since recent studies confirm that SNe~IIL have somewhat
higher explosion energies per unit ejecta mass than SNe~IIP 
\citep[see, for example,][]{faran:14b,gall:15} and the
steeper decline of SNe~IIL is due to a larger amount of
dense CSM \citep{morozova:17}. The important
new thing that we find here is that the correlation between $M_{\rm CSM}$
and explosion energy seems to extend continuously and roughly linearly
along the entire sample. What this actually means for the mechanism
that generates the dense CSM is unclear. One possibility is that whatever
process that generates the CSM also decreases the gravitational binding
energy of the star. Then the mechanism that unbinds the star, commonly
presumed to be neutrino heating, is more able to create a more energetic
explosion.

\begin{figure}
  \centering
  \includegraphics[width=0.475\textwidth]{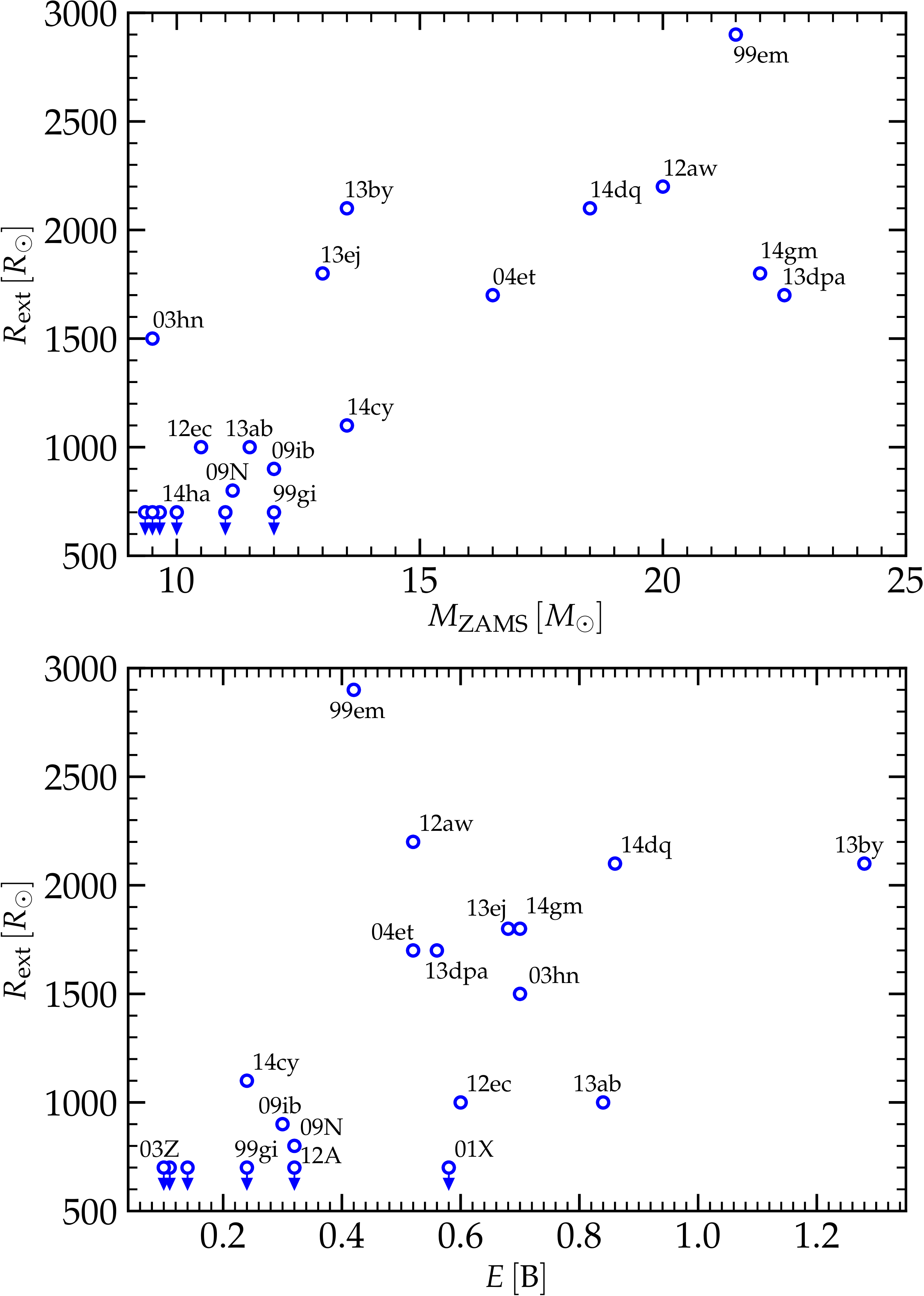}
  \caption{CSM radius versus ZAMS mass (top panel) and
  explosion energy (bottom panel), obtained from the 
  numerical light curve fits. Arrows indicate SNe for which we only
  have upper limits on CSM mass from Table~\ref{tab:SN2}.} 
  \label{fig:CSM_rad}
\end{figure}

Figure~\ref{fig:CSM_rad} shows the inferred radii of the CSM versus
the ZAMS masses of the progenitors (top panel) and the explosion 
energies (bottom panel). There appears to be no clear correlation
between $R_{\rm ext}$ and the explosion energy. On the other hands,
there may be indication that the dense CSM is more extended for
larger $M_{\rm CSM}$.
This radius is related to both the duration
and velocity of the CSM.
Generally speaking, these values $R_{\rm ext}$ are only a few
 times larger than the progenitors' radii. This indicates that no matter
 the velocity, the dense CSM must have been generated soon before the
 explosion. For example, assuming the velocity
of $10\,{\rm km}\,{\rm s}^{-1}$ gives a timescale of just years
(and maybe just months for higher velocities). The advanced stages
of nuclear burning could act on such timescales, but the key issue
is whether there is a way to harness the energy from this burning to
generate the dense CSM.
There have been a number of theoretical studies on how to generate
pre-explosion outbursts \citet{yoon:10,arnett:11,quataert:12,shiode:14,smith:14,moriya:15,
woosley:15,quataert:16}, and most recently one of the most detailed studied
of outbursts was modeled by \citet{fuller:17}
using open source
stellar evolution code \texttt{MESA} \citep{paxton:11,paxton:13,paxton:15}.
Nevertheless, it has yet to be shown that these mechanisms can exactly
reproduce the dense CSM properties as we need here.

\subsection{Other implications of dense CSM}

A number of observational works on the early SN~II light curves
point out that they typically rise faster than the analytical 
and hydrodynamical light curves obtained from standard RSG
models \citep[see, for example,][]{gall:15,gonzalez:15}. 
As a consequence, fitting the early SN~II light curves
with the analytical and numerical models resulted in 
relatively small 
($\sim$500$\,R_{\odot}$) estimated radii of their progenitors 
(\citealt{dessart:13,gonzalez:15,garnavich:16}; see, 
however, \citealt{shussman:16,rubin:16a}). 
These values are on the lower end of the radii estimated from the
observations of galactic and Magellanic Clouds RSGs by \citet{levesque:05}
and \citet{levesque:06}, which lie in the range between $\sim$400
and $\sim$1500$\,R_{\odot}$. At the same time, a more recent study
by \citet{davies:13} suggests that the previous works could underestimate the RSG 
temperatures and, consequently, overestimate their radii. For comparison,
the radii of the \texttt{KEPLER} models we use vary between $\sim$400
and $\sim$1400$\,R_{\odot}$ without the CSM, and similar values for the models
with the same ZAMS masses are obtained in \citet{renzo:17}. Our current study 
shows that adding dense CSM to otherwise standard RSGs
leads to faster rise times, 
even though the total radius of the models
increases. As can be seen from Figures~\ref{fig:fits1} 
and~\ref{fig:fits2}, this effect is weaker for 
the ultraviolet bands, which can have the same rise time with and
without CSM, but becomes very strong in red bands, where it is
crucial for reproducing the early rise and maximum. Therefore,
CSM may serve as an explanation of the fast rise of the early 
SN~II light curves rather than the common explanation of changing the radius of the 
underlying RSGs. 

Although we focus on photometric light curves,
the presence of a dense CSM around otherwise normal SNe~IIP has
also been recently suggested by the spectroscopic observations of a number
of events \citep[see, for example,][]{quimby:07,kiewe:12,yaron:17}. 
The key is to obtain the spectra early enough, since
the characteristic lines typically disappear within a day after the explosion.
The material probed by these observations is smaller in mass and larger
in radius than what we study here \citep[see discussions in][]{yaron:17,dessart:17},
so the exact relation between these two components of CSM is not clear.
It is important to note that in our models the shock breakout
happens at the very outer edge of the CSM. This means that at the moment
when the first SN light is seen, the bulk of this CSM is 
optically thick and cannot contribute to the early spectrum. 
In addition, \citet{dessart:17} have shown that 
a sharp transition between the low and high mass loss states assumed
in our models cannot reproduce the narrow emission lines observed, for example, 
in SN 2013fs. In reality, it is likely that there is a smoother transition between the 
dense CSM we model in this work and the regular low
density RSG wind. One interesting possibility to explain it is by employing
an accelerating wind that goes
from being dense close to the star and less dense as it accelerates further
from the star \citep{moriya:17}. In this case, the shock breaks out
inside the wind, which changes the timescale of the breakout from the expected hours to
days \citep{moriya:11}. Such an extended shock breakout was seen, for example, in early
observations of a Type II SN PS1-13arp by \citet{gezari:15}.

\subsection{Limitations of \texttt{SNEC}}

There are a number of approximations and simplifications made
in the numerical code we are using, so it is natural to ask
how these impact our inferences of dense CSM.
Generally speaking, comparison between the light curves generated by \texttt{SNEC} 
and the multi-group radiation-hydrodynamic code \texttt{STELLA}
\citep{blinnikov:93,baklanov:05,blinnikov:06,blinnikov:11,kozyreva:17}
that uses the same RSG progenitor shows good agreement
in the bolometric luminosity and plateau duration (P.~Baklanov, 
S.~Blinnikov, private communication). Nevertheless, one of the largest discrepancies
between the color light curves is seen during the early rise, where 
\texttt{SNEC} light curves rise consistently faster than \texttt{STELLA}
light curves, reaching the same magnitudes $\sim3-5$ days earlier
\footnote{The largest disagreement is seen in the $U$- and $B$-band
light curves, which drop faster in \texttt{STELLA}
simulations after day $\sim20$ due to the proper treatment 
of the iron line opacity. Since we are
aware of this problem in our simulations, which is the main reason we do not fit the 
observed light curves in $U$- and $B$-bands, we omit the discussion of this effect
in the text.}. This can be explained by incomplete thermal equilibrium 
between the radiation and matter at the photosphere, which is captured 
by \texttt{STELLA} but not by \texttt{SNEC} (the current version of 
\texttt{SNEC} uses the assumption of local thermodynamical equilibrium
across the entire model). In the more realistic case, the observed temperature is
set in a region deeper than the photosphere, 
where radiation processes are
able to couple the gas and radiation \citep{nakar:10}. The value of
the observed temperature is therefore higher, and it takes longer for it to
drop down to the values corresponding to the black body maxima in the
optical bands. This leads to the slower rise in these bands as found by
\texttt{STELLA}. As a consequence, we expect that it would be even more challenging 
for the multi-group radiation-hydrodynamic codes to reproduce
the observed SN~IIP light curves using stellar evolution RSG 
models without invoking some sort of CSM. Therefore, the mass of
the CSM needed to fit the light curves with these codes will be probably even
larger than our estimates. This is roughly consistent with our modeling of
the SN 2013fs where 
the CSM mass is estimated to be $0.47\,M_{\odot}$ \citep{morozova:17}
and  the work of \citet{moriya:17} where it is estimated to be 
$0.5\,M_{\odot}$.

Another limitation is related to the stiff inner boundary in \texttt{SNEC},
which causes numerical problems when we explode high ZAMS mass progenitors
with low explosion energies. In Figure~\ref{fig:chisq1}, the SNe that are most strongly
affected by this issue are 1999em, lsq13dpa, 2014cy and asassn14ha. In order to check
how this alters the inference of minimum and maximum ZAMS masses of
Type II progenitors, we excluded these four SNe from the set and repeated the analysis 
shown in Figure~\ref{fig:likelihood}. Without these SNe, the minimum mass $M_{\rm ZAMS,\ min}$
does not change, while the maximum mass $M_{\rm ZAMS,\ max}$ changes only slightly from 
$22.9\,M_{\odot}$ to $22.5\,M_{\odot}$. Therefore, we believe that the main conclusions of
our study do not depend on this problem.


\section{Conclusions}
\label{conclusions}

Using modeling of the multi-band light curves of twenty SNe II, we
have for the first time constrained the
progenitor ZAMS mass,
 explosion energy, and the mass and radial extent of dense CSM.
For 25\% of the SNe
we infer  ZAMS masses larger than $\sim17\,M_{\odot}$, the maximum
limit suggested by the pre-explosion imaging. We show that the
mass distribution
we infer for our sample is consistent with a Salpeter distribution
with a minimum and 
maximum ZAMS masses for the SNe~II progenitors are equal to 
$10.4$ and $22.9\,M_{\odot}$, respectively. This is in rough agreement with the
observed masses of RSGs and suggests a solution to the RSG problem.

Our results imply that dense CSM is very common amongst SNe~IIP,
at least $70\%$ of our fits benefit substantially from including it in the
model. The largest amount of CSM (in mass) is expected from IIL-like
events, while underluminous SNe from the low mass progenitors have
the smallest amount. The radii of the CSM that we deduce are quite small,
in a range of $\sim800-3000\,R_{\odot}$, which argues that whatever mechanism
generates the CSM occurs years if not months before the explosion.
Given these short timescales, the formation of CSM may be related to the advanced stages of nuclear 
burning in the stellar interior, which have recently been studied as a mechanism
for generating mass outbursts \citep[see][]{quataert:12,shiode:14,
woosley:15,quataert:16,fuller:17}.

With this sample we are able to explore whether the properties of the dense
CSM are related to other aspects of the progenitor or explosion. The main
correlations we find are that larger ZAMS masses have CSM at larger radii
and larger energy explosions have larger mass CSM.
In the former case, this
may be related to the time when the dense CSM is generated. In the latter
case, there is a suggestion that whatever generates the dense CSM may
also make the progenitor star easier to explode.
There is of course a many decades long history of trying to understand how core-collapse
SNe explode by way of the neutrino mechanism with more failure than
success. One possible solution is that whatever mechanism generates
the dense CSM may also change the structure of the star in a critical
way that could help make the star easier to unbind. Such a hypothesis will hopefully
be explored in future theoretical work by utilizing non-standard RSG structures (motivated
by the need to generate dense CSM) in
the most sophisticated calculations that investigate the neutrino mechanism.

\acknowledgments
We acknowledge helpful discussions with and feedback from A.~Burrows, N.~Smith,
J.~Stone and D.~Radice.
We thank P. Baklanov, S. Blinnikov, C. Ott, E. Sorokina and T. Moriya
for comparison between \texttt{SNEC} and \texttt{STELLA} light curves. 
Computations were performed on the TIGER cluster at Princeton
University. V.~M. acknowledges funding support from the Lyman 
Spitzer Professorship at Princeton University and NSF Grant AST-1714267.
A.L.P. acknowledges financial support for this research from a Scialog 
award made by the Research Corporation for Science Advancement.

\bibliographystyle{apj}

\end{document}